\documentclass[12pt,preprint]{aastex}

\usepackage{color}
\usepackage{graphicx}

\slugcomment{Accepted to ApJ}

\shorttitle{Chemical Abundances in a Sample of Red Giants in the Open Cluster NGC 2420 from APOGEE}
\shortauthors{Souto et al.}

\begin{document}


\title{Chemical Abundances in a Sample of Red Giants in the Open Cluster NGC 2420 from APOGEE}

\author{Diogo Souto\altaffilmark{1}}
\email{souto@on.br}
\author{K. Cunha\altaffilmark{1}}
\author{V. Smith\altaffilmark{2}}
\author{C. Allende Prieto\altaffilmark{3,4}}
\author{M. Pinsonneault\altaffilmark{5}}
\author{O. Zamora\altaffilmark{3,4}}
\author{D. A. Garc\'ia-Hern\'andez\altaffilmark{3,4}}
\author{Sz. M\'esz\'aros\altaffilmark{6}}
\author{J. Bovy\altaffilmark{7,8}}
\author{A. E. Garc\'ia P\'erez\altaffilmark{3,4}}
\author{F. Anders\altaffilmark{9}}
\author{D. Bizyaev\altaffilmark{10,11}}
\author{R. Carrera\altaffilmark{3,4}}
\author{P. M. Frinchaboy\altaffilmark{12}}
\author{J. Holtzman\altaffilmark{13}}
\author{I. Ivans\altaffilmark{14}}
\author{S. R. Majewski\altaffilmark{15}}
\author{M. Shetrone\altaffilmark{16}}
\author{J. Sobeck\altaffilmark{15}}
\author{K. Pan\altaffilmark{10}}
\author{B. Tang\altaffilmark{17}}
\author{S. Villanova\altaffilmark{17}}
\author{D. Geisler\altaffilmark{17}}

\altaffiltext{1}{Observat\'orio Nacional, Rua General Jos\'e Cristino, 77, 20921-400 S\~ao Crist\'ov\~ao, Rio de Janeiro, RJ, Brazil}

\altaffiltext{2}{National Optical Astronomy Observatory, 950 North Cherry Avenue, Tucson, AZ 85719, USA}
\altaffiltext{3}{Instituto de Astrof\'isica de Canarias (IAC), V\'ia Lactea S/N, E-38205, La Laguna, Tenerife, Spain}
\altaffiltext{4}{Departamento de Astrof\'isica, Universidad de La Laguna (ULL), E-38206, La Laguna, Tenerife, Spain}
\altaffiltext{5}{Department of Astronomy, The Ohio State University, Columbus, OH 43210, USA}

\altaffiltext{6}{ELTE Gothard Astrophysical Observatory, H-9704 Szombathely, Szent Imre Herceg st. 112, Hungary}

\altaffiltext{7}{Department of Astronomy and Astrophysics, University of Toronto, 50 St. George Street, Toronto, ON, M5S 3H4, Canada}

\altaffiltext{8}{Dunlap Institute for Astronomy and Astrophysics, University of Toronto, ON M5S 3H4, Canada}

\altaffiltext{9}{Leibniz-Institut für Astrophysik Potsdam (AIP), An der Sternwarte 16, 14482 Potsdam, Germany}

\altaffiltext{10}{Apache Point Observatory and New Mexico State University, P.O. Box 59, Sunspot, NM, 88349-0059, USA}

\altaffiltext{11}{Sternberg Astronomical Institute, Moscow State University, Moscow}

\altaffiltext{12}{Department of Physics and Astronomy, Texas Christian University, Fort
Worth, TX 76129, USA}

\altaffiltext{13}{New Mexico State University, Las Cruces, NM 88003, USA} 

\altaffiltext{14}{Department of Physics and Astronomy, The University of
Utah, Salt Lake City, UT 84112, USA}

\altaffiltext{15}{Department of Astronomy, University of Virginia, Charlottesville, VA 22904-4325, USA}

\altaffiltext{16}{University of Texas at Austin, McDonald Observatory, USA}

\altaffiltext{17}{Departamento de Astronom\'ia, Casilla, 160-C, Universidad de Concepcion, Concepcion, Chile}

\begin{abstract}
NGC 2420 is a $\sim$2 Gyr-old well-populated open cluster that lies about 2 kpc beyond the solar circle, in the general direction of the Galactic anti-center. Most previous abundance studies have found this
cluster to be mildly metal-poor, but with a large scatter in the
obtained metallicities for this open cluster. Detailed
chemical abundance distributions are derived for 12
red-giant members of NGC 2420 via a manual abundance analysis of 
high-resolution (R = 22,500) near-infrared 
($\lambda$1.5 - 1.7$\mu$m) spectra obtained from the Apache Point 
Observatory Galactic Evolution Experiment (APOGEE) survey.  
The sample analyzed contains 6 stars that are identified as members of the first-ascent red giant branch (RGB), as well as 6 members of the red clump (RC). 
We find small scatter in the star-to-star abundances in NGC 2420, with a mean cluster abundance of [Fe/H] = -0.16 $\pm$ 0.04 for the 12 red giants.
The internal abundance dispersion for all elements
(C, N, O, Na, Mg, Al, Si, K, Ca, Ti, V, Cr, Mn, Co and Ni) is also
very small ($\sim$0.03 - 0.06 dex), indicating a uniform cluster abundance distribution within the uncertainties. NGC 2420 is one of the clusters used to calibrate the APOGEE
Stellar Parameter and Chemical Abundance Pipeline (ASPCAP). The results from this manual analysis compare well with ASPCAP abundances for most of the elements studied, although for Na, Al and V there are more significant offsets. No evidence of extra-mixing at the RGB luminosity bump is found in the $^{12}$C and $^{14}$N abundances from the pre-luminosity-bump RGB stars in comparison to the post-He core-flash RC stars. 

\end{abstract}


\section{Introduction}

The open cluster NGC 2420, with an age of roughly 2 Gyr, is 
located towards the Galactic anti-center at a Galactocentric distance of 10.78 kpc (Sharma et al. 2006). Given its age, location, and metallicity, this cluster is an interesting object for studies of Galactic chemical evolution. The first detailed photometric study of NGC 2420 was by Sarma $\&$ Walker (1962) and, later, West (1967) noted its stars exhibited a mild excess in $\delta$(U - B), which suggested the cluster was somewhat metal-poor.
The earliest determinations of spectroscopic metallicities for NGC 2420 were made by Pilachowski et al. (1980), Cohen (1980), and Smith \& Suntzeff (1987); these studies found a small range of metallicities clustering around [Fe/H] $\approx$ -0.60.  Later studies using photometric data and isochrones (Anthony-Twarog et al. 2006) derived somewhat higher values of [Fe/H] $\approx$ -0.30. More recently, the high-resolution spectroscopic study by Pancino et al. (2010) found NGC 2420 to be considerably more metal-rich, with [Fe/H] $\approx$ -0.05 dex. Meanwhile Jacobson et al. (2011) analyzed spectra of moderately high-resolution (R $\approx$ 18,000) and found an average metallicity for this cluster of [Fe/H] $\approx$ -0.20 dex. The large scatter for [Fe/H] in the literature suggests that a new abundance analysis using different spectra would be worthwhile and here a sample of red giant members of NGC 2420 are analyzed using near-infrared (NIR) high-resolution spectra from the SDSS-III/APOGEE survey (Apache Point Observatory Galactic Evolution Experiment; Einsenstein et al. 2010; Majewski et al. 2016).

The APOGEE-1 survey observed more than 146,000 Galactic red-giants in three years of operation having ended in July 2014. A number of red-giants in disk open clusters, including NGC 2420, were targeted by APOGEE-1 to serve as calibration clusters for the survey, to study cluster membership, and measure Galactic metallicity gradients. Stellar parameters (effective temperatures and surface gravities), chemical abundances of several elements, and metallicities for all the stars observed in the APOGEE survey are derived automatically by means of the pipeline ASPCAP (APOGEE Stellar Parameter and Chemical Abundances Pipeline; Garc\'ia P\'erez et al. 2016). 

This paper presents chemical abundances for 12 red-giant members of the open cluster NGC 2420 using a manual spectroscopic chemical abundance analysis in the same way as made by Cunha et al. (2015) and Smith et al. (2013). We derive stellar parameters and the abundances of 16 elements: C, N, O, Na, Mg, Al, Si, K, Ca, Ti, V, Cr, Mn, Fe, Co and Ni. One of the goals of this study is to provide a direct comparison of the results from a manual abundance analysis with those derived automatically by the ASPCAP pipeline. The APOGEE team is continually improving ASPCAP and the most recent version of ASPCAP has produced the stellar parameters and metallicity results for the 13$^{th}$ SDSS Data Release, hereafter, DR13, which will become publicly available in summer 2016. The results presented in this independent work will help to verify ASPCAP. 

\section{The APOGEE Spectra}

APOGEE spectra are obtained with a 300-fiber cryogenic spectrograph on the 2.5m Telescope at the Apache Point Observatory obtaining high-resolution spectra (R $\sim$ 22,500) between $\sim$ $\lambda$1.5--1.7 $\mu$m (Wilson et al. 2010, Gunn et al. 2006). The reduction of the APOGEE spectra, as well as the determination of radial velocities was carried out by the data reduction pipeline (Nidever et al. 2015) using reduction scripts designed for DR13. Each spectrum analyzed here was combined from multiple visits, typically 3 to 5, resulting in very high signal-to-noise ratio spectra (S/N $>$ 100) for all targets. 

APOGEE-1 targeted 19 red-giants (Zasowski et al. 2013) as possible members of NGC 2420 (these are labeled as ${\it APOGEE}$ ${\it calibration}$ ${\it cluster}$ in the DR13 tables; Frinchaboy et al. 2013). However, it was found that 7 of these targets are either binaries or not members of the cluster given their inconsistent radial velocities when compared to that expected for this open cluster of $\sim$ +73 Km/s (Smith $\&$ Suntzeff 1987; Liu $\&$ James 1987). The final sample of red giants studied here, their measured radial velocities and dispersion from individual visits, as well as the signal-to-noise ratios for the combined spectra are found in Table 1.

\section{Determination of Effective Temperatures and Surface Gravities}

Standard stellar chemical abundance analysis requires a pre-determined set of atmospheric parameters -- effective temperature (T$_{\rm eff}$), surface gravity, and metallicity, that are used to compute model atmospheres. In this analysis, photometric calibrations were used to derive effective temperatures, while stellar mass and luminosity were used to calculate the surface gravities, and Fe I lines to derive both microturbulent velocities and metallicities. The adopted atmospheric parameters for the studied stars are presented in Table 1.

The effective temperatures were obtained using the photometric calibrations of Gonz\'alez-Hern\'andez $\&$ Bonifacio (2009) for the colors $V$-$J$, $V$-$H$, $V$-$K$$_{\rm S}$ and $J$-$K$$_{\rm S}$. The $J$, $H$ and $K$$_{\rm S}$ magnitudes are from 2MASS and $V$ magnitudes from UCAC4 (Zacharias et al. 2013), NOMAD (Zacharias et al. 2005) and Anthony-Twarog et al. (2006).  Relations in Schlegel et al. (1998) and Carpenter (2001) were used to derive the de-reddened colors, using a reddening of E(B-V) = 0.05 (Salaris et al. 2004; Grocholski $\&$ Sarajedini 2003; Anthony-Twarog et al. 2006). Figure 1 shows the effective temperature calibrations corresponding to the four different colors considered; the final effective temperatures adopted for the stars (shown as red points) are the mean of the T$_{\rm eff}$ obtained for each color. The errorbars represent the standard deviation of the mean and typical values are $\pm$ 50K. As an estimate of the sensitivity of the effective temperatures to the adopted reddening, an extreme change in the reddening by 0.05 magnitudes would impart a difference in the derived effective temperatures of $\approx$ 120K, averaged over all colors. 

The surface gravities were determined from fundamental relations (Eq. 1). Stellar masses of M$_{\star}$ $\sim$ 1.6 M$_{\odot}$ were estimated using PARSEC isochrones (Bressan et al. 2012) for a cluster age of 2 Gyr (Sharma et al. 2006) and [M/H] = -0.20 dex. Absolute magnitudes were derived using the distance modulus of (m - M)$_{0}$ = 11.94 (Salaris et al. 2004) and bolometric corrections from Montegriffo et al. (1998). 
The solar values adopted were: log g$_{\odot}$ = 4.437 dex, T$_{\rm eff, \odot}$ = 5770 K and M$_{bol,\odot}$ = 4.75 (Andersen 1999). 

\begin{equation}
\log{g} = \log{g_{\odot}} + log\left(\frac{M_{\star}}{M_{\odot}}\right) + 4log\left(\frac{T_{\star}}{T_{\odot}}\right) + 0.4(M_{bol,\star} - M_{bol,\odot}),
\end{equation}

Figure 2 (top panel) shows the selected isochrone in the effective temperature versus H$_{0}$ magnitude plane. The derived T$_{\rm eff}$ for ten of the targets cluster around T$_{\rm eff}$ = 4901 $\pm$ 63 K; one star deviates clearly from this group as it is much cooler (T$_{\rm eff}$ $\sim$ 4209 K) and further up the red-giant branch, while another is just slightly hotter (T$_{\rm eff}$ $\sim$ 5110 K). In this diagram, the $H$$_{0}$ magnitudes can be used to identify the slightly more luminous clump giants (filled circles) and less luminous first-ascent red giant branch stars (open circles), as defined by the isochrones. We recognize that the same six selected RC stars segregate in the spectroscopic HR diagram shown in the bottom panel of Figure 2, where log g is plotted versus T$_{\rm eff}$. 

The bottom panel of Figure 2 takes the physical values of log g calculated from Equation 1 and plots
them versus photometric T$_{\rm effs}$, with the selected isochrone overplotted. The agreement
between the isochrone and derived stellar parameters is very good.  The segregation between the
RGB and RC stars noted for their values of $H$$_{0}$ carries over into a clear segregation in
log g; the difference in log g between an RGB and RC star in NGC 2420 is about 0.25 dex for a
given T$_{\rm eff}$.

The small scatter found in the observed red giants about the RGB and RC isochrone suggests
that the internal values of log g have small uncertainties, with $\Delta$log g $\le$ 0.05 dex.
Systematic offsets in log g are undoubtedly somewhat larger, due to uncertainties in the assumed turnoff mass,
caused by uncertainties in the cluster age, as well as errors in the distance.  A range of
cluster ages have been estimated for NGC 2420: 2 Gyr (Sharma et al. 2006), 3.4 Gyr (Anthony-Twarog
et al. 1990), and 4.0 Gyr (McClure et al. 1978).  The scatter within these age estimates is
1.0 Gyr, which would cause offsets in the log g scale of about $\pm$0.10 dex.

\section{Abundance Analysis}
A total of 70 spectral features were analyzed for the computation of the chemical abundances of 16 elements: C, N, O, Na, Mg, Al, Si, K, Ca, Ti, V, Cr, Mn, Fe, Co and Ni. A classical LTE manual abundance analysis of the APOGEE spectra was performed with the code MOOG (Sneden 1973) using spectrum synthesis. The model atmospheres used were calculated for the APOGEE-1 project by M\'esz\'aros et al. (2012) and these are one-dimesional plane parallel models from the Kurucz ATLAS9 grid (Kurucz 1993). 

The line list adopted for the calculation of the synthetic spectra was developed by the APOGEE/ASPCAP team and it is part of DR13 (this line list designation is 20150714). 
The details concerning the construction of the APOGEE line list (for DR12) can be found in Shetrone et al. (2015), and the changes in DR13 will be presented elsewhere.

Examples of synthetic fits to a portion of the observed APOGEE spectra (spectral region covering 15600 to 15700 \AA{}) are shown in Figure 3 for three sample giants with effective temperatures spanning the range of our sample: T$_{\rm eff}$ $\sim$ 4200 -- 5100 K. Those lines used in the abundance measurements are indicated in the figure. The best fitting spectra were selected visually and the quality of the fits presented are typical of what was obtained for other regions of the APOGEE spectra. 
Red-giants have typically low rotational velocities ($\le$8 km-s$^{-1}$; e.g. Carlberg et al. 2016), along with macroturbulent velocities of $\sim$7 km-s$^{-1}$ (Grey 1978). The spectra were fit using only a Gaussian profile with a broadening corresponding to a full width half maximum (FWHM) of $\sim$ 730 m\AA{}, or $\sim$ 13.7 km/s in velocity broadening. We tested different values of stellar V sin$(i)$, as well as macroturbulent velocities, using the broadening tool in MOOG, but did not detect any excess broadening beyond the FWHM of the instrumental profile corresponding to the APOGEE resolution. 

\subsection{Metallicities and Microturbulent Velocities}

Iron is often used as a proxy for the overall stellar metallicity and nine Fe I lines (Table 3) were used to set the Fe abundance in this study.
The sample Fe I lines were used to derive the microturbulent velocities ($\xi$), in a manner similar to the previous work by Smith et al. (2013). Iron abundances were derived for different values of $\xi$ and the selected microturbulence was the one that produced the minimum spread in the Fe I abundances. The adopted values of microturbulent velocities for the stars can be found in Table 1.

\subsection{C, N, O and Other Elements}

Carbon, nitrogen, and oxygen are key elements being probed by APOGEE (Majewski et al. 2016). Their abundances are important in studying nucleosynthesis and chemical evolution, as well as mixing and dredge-up in red-giant stars. The APOGEE spectral window contains numerous CO, OH, and CN lines and is well-suited for determining C, N, and O abundances.

We use the CO lines to derive the carbon abundances, lines of CN for nitrogen and the OH lines to obtain oxygen abundances. The molecular lines discussed here are composed of the dominant isotopes of each element and, thus, consist of $^{12}$C$^{16}$O, $^{12}$C$^{14}$N, and $^{16}$OH.  The same methodology as described in Smith et al. (2013) was used to obtain a solution for C, N and O abundances that satisfies the fitting of all molecular lines consistently: we first derive carbon abundances from CO, then derive oxygen from OH and nitrogen from CN lines. The molecular transitions and spectral regions used in our analysis are listed in Table 2. This set of lines/regions is the same as in Smith et al. (2013) and these were adequate for the analysis of our target stars, which are mostly hotter than those in Smith et al. (2013). We note, however, that the region containing the $^{12}$C$^{16}$O lines from the (3-0) vibration-rotation transitions covering the wavelength between 15578 -- 15586\AA\ become weakly dependent on the carbon abundance and varied mostly with the nitrogen abundance in the range of T$_{\rm eff}$ $\sim$ 4800K. These features were then used in conjunction with other lines of CN to help constrain the nitrogen abundances. 

The APOGEE spectra contain several lines arising from atomic transitions of a number of elements produced in most major nucleosynthetic sites. These include spectral lines of the alpha-elements such as: Mg I, Si I, Ca I and Ti I; spectral lines of the odd-Z elements: Na I, Al I, K I, as well as Fe-peak elements, such as: V I, Cr I, Mn I, Fe I, Co I and Ni I. The atomic lines analyzed in this study and the individual abundances measured in each case are listed in Table 3.

\subsection{Abundance Sensitivies}
 
Table 4 presents the sensitivity of the derived abundances to changes in the effective temperature, surface gravity, microturbulent velocity and metallicity, similarly to the discussion in Smith et al. (2013). We adopted a base model that is representative of our red-giant sample: T$_{\rm eff}$ = 4900 K, log g = 2.70, [M/H] = -0.20. For the manual analysis conducted here, abundance uncertainties were calculated in a simplified way by varying each stellar parameter individually and computing a new model atmosphere.  In these tests, the effective temperature was changed by 50K, the surface gravity value by 0.20 dex and the metallicity by 0.20 dex. The microturbulent velocity adopted was 1.40 Km.s$^{-1}$ and it was varied by 0.20 Km.s$^{-1}$.
The changes in stellar parameters are not very different from typically expected uncertainties for this type of spectroscopic analysis, although the perturbation in metallicity is somewhat conservative and larger than the expected uncertainty. 

A quadrature sum of the stellar parameter errors from Table 4 reveals uncertainties that are $\le$ 0.10 dex, except for Ti I and O, which present the largest uncertainties: $\Delta$A(Ti) = 0.10 dex and $\Delta$A(O) = 0.13 dex. The Ti abundances (from Ti I) are dominated by sensitivity to T$_{\rm eff}$, while the O abundances (from OH) are dominated by sensitivity to overall model metallicity.  

The observed dispersions in the line-to-line abundances are presented in Tables 2 and 3 (the mean abundances and standard deviations of the mean are found below the individual line abundances for each studied element). The abundances derived from the different lines are overall quite consistent with sigmas typically around $\pm$0.04 dex.

This level of dispersion can certainly be accounted for by the uncertainties in the gf-values, as well as modest errors in the stellar parameters of the approximate magnitude as investigated in Table 4. The elements that exhibit larger line-to-line scatter in their abundances are magnesium (six Mg I lines measured) and silicon (eight Si I lines measured): $\langle$$\sigma$$\rangle$ = 0.09 dex and $\langle$$\sigma$$\rangle$ = 0.07 dex, respectively. It should be noted that the Mg I and Si I lines are some of the stronger lines in the APOGEE spectral window. Three elements have only one well-defined spectral line each in the APOGEE region and the abundances of these species should be treated with more caution: V I, Cr I and Co I. 

\section{Discussion}
 
We analyzed 12 red giant members of NGC 2420: six from the red clump and six from the red-giant branch. Line-by-line measurements of the iron abundances for all studied stars are presented in Table 3; the individual elemental abundances have typical  standard deviations of the mean that are less than 0.07 dex. There is also small scatter in the star-to-star abundances in NGC 2420, with a mean cluster abundance and standard deviation of the mean of $\langle$A(Fe)$\rangle$ = 7.29 $\pm$ 0.04 for the 12 giants. This translates to $\langle$[Fe/H]$\rangle$ = -0.16 $\pm$ 0.04 for NGC 2420 by using Asplund et al. (2005) as Solar reference. The mean C and N abundances obtained for the stars in our sample are quite consistent and indicate a small standard deviation of the mean values: $\langle$[C/Fe]$\rangle$ = -0.07 $\pm$ 0.04, $\langle$[N/Fe]$\rangle$ = +0.17 $\pm$ 0.03. These carbon and nitrogen results are overall consistent with the CN-Cycle, given that the abundance of carbon is down (slightly below the solar scaled value) and the abundance of nitrogen is enhanced relative to the solar scaled value (Section 5.2).

The alpha element oxygen is also mildly enhanced: $\langle$[O/Fe]$\rangle$ = +0.10 $\pm$ 0.03. We note that this spread is very similar to the values found by Bertran de Lis et al. (2016) for stars with similar temperatures in other clusters with metallicities near solar, such as M67, NGC 6819 and NGC 2158. The mean abundances for the other alpha elements, however, are roughly solar scaled with the mean value for Mg, Si, Ca and Ti being $\langle$[$\alpha$/Fe]$\rangle$ = $\langle$[(Mg+Si+Ca+Ti/4)/Fe]$\rangle$ = +0.01 $\pm$ 0.02 dex. For the iron peak elements we obtained: $\langle$[(Cr+Mn+Co+Ni/4)/Fe]$\rangle$ = -0.06 $\pm$ 0.02 dex, while the odd-Z elements Na, Al and K show a marginal enhancement of $\langle$[(Na+Al+K/3)/Fe]$\rangle$ = +0.06 $\pm$ 0.06 dex. 

\subsection{Comparisons with ASPCAP and the Literature}

One of the objectives of this study is to compare the results from the APOGEE automated abundance analysis derived using ASPCAP with an independent manual abundance analysis. ASPCAP abundances and stellar parameters are obtained from automatic matches of APOGEE spectra to synthetic libraries (Zamora et al. 2015) for a 6- or 7-D optimization of T$_{\rm eff}$, logg, [M/H], [C/Fe], [N/Fe], [$\alpha$/Fe] and sometimes ($\xi$) using the FERRE code (Allende Prieto et al. 2006). DR13 includes both raw ASPCAP values, as well as calibrated values that were adjusted in order to match literature abundances of selected calibrators (see discussion in Holtzman et al. 2015).

\subsubsection{Stellar Parameters}

Figure 4 shows an H-R diagram plotted as log g versus T$_{\rm eff}$ for the target stars. DR13 results for both raw and calibrated ASPCAP abundances are also shown. This comparison indicates that there is a clear offset between the stellar parameters derived in this study (red circles) and the raw values from ASPCAP (brown pentagons), while the calibrated ASPCAP values (grey diamonds) show overall much better agreement with our results. It can be seen from the top left panel of Figure 5 that our effective temperatures (computed from photometric calibrations; Section 3) agree quite well with the ASPCAP T$_{\rm effs}$, which are derived purely from the APOGEE spectra. There is just a small tendency for our effective temperatures to be hotter than those from ASPCAP: the average difference between the two independent scales is $\langle$$\delta$(T$_{\rm eff}$(This work - ASPCAP)$\rangle$ = 49 $\pm$ 22 K. (We note that ASPCAP effective temperatures were not calibrated for DR13). We also show in the bottom left panel the T$_{\rm eff}$ results from Jacobson et al. (2011) and Pancino et al. (2010) for a sample of stars that we have in common with those studies (Table 1). 
The effective temperatures from Jacobson et al. (2011; green triangles) and Pancino et al. (2010; blue squares), which are both derived from the photometric calibrations in Alonso et al. (1999), do not show significant offsets with our results. 

The surface gravity comparisons are shown in the right panels of Figure 5. Our derived log g values agree very well with those obtained by Pancino et al. (2010; blue squares) and Jacobson et al. (2011; green triangles) for the stars in common. This is expected because those previous log g derivations are based on physical relations (Eq. 1). It is also clear from this figure that the surface gravity results in DR13, which come directly from the ASPCAP analysis of the APOGEE spectra (brown pentagons), are systematically larger than the log g values obtained from fundamental relations: 
$\langle$$\delta$(log g(This work - ASPCAP)$\rangle$ = -0.26 $\pm$ 0.12. We note that for the RC sample the log g difference is $\delta$ = -0.34 $\pm$ 0.10 while for the RGB sample $\delta$ = -0.18 $\pm$ 0.07. 

This systematic offset in the ASPCAP derived surface gravities was also noticed in the previous APOGEE data releases (DR10, Ahn et al. 2014 and DR12, Alam et al.2015) and calibrations have been applied to correct for this bias (see discussions in Holtzman et al. 2015 and M\'esz\'aros et al. 2013). The calibration of the ASPCAP log g results in DR13 uses an algorithm for deciding if a star is on the RC or RGB based on its T$_{\rm eff}$, log g and [C/N] abundances.
DR13 ASPCAP calibrated log g values show, on average, much better agreement with our log g (non-spectroscopic) determinations: $\langle$$\delta$(log g(This work - ASPCAP$_{\rm calibrated}$)$\rangle$ = 0.00 $\pm$ 0.12. The source of the offset between the uncalibrated ASPCAP values of log g and the physical log g's is unknown and we note that the APOGEE spectra themselves cannot be used the Fe I/Fe II ionization balance as no Fe II lines are detected in APOGEE spectra.

\subsubsection{Chemical Abundances}

Elemental abundances obtained for the NGC 2420 stars, along with the raw and calibrated ASPCAP results, are shown in Figure 6 as a function of the effective temperatures derived here. The calculated mean abundance differences between our results and ASPCAP are also indicated in each panel of Figure 6. 

For a significant fraction of the elements, the abundances obtained manually are similar to those derived automatically by ASPCAP, with all 3 types of results (manual, ASPCAP raw, and ASPCAP calibrated) agreeing in the mean to $\sim$0.05 dex. This is the case for the elements: C, Mg, K, Ca, Cr, Mn, Fe, and Ni.  

The remaining 8 elements exhibit offsets between the mean abundances of these three sets which are greater than $\sim$0.05 dex.  In the case of O and Al, in particular, the ASPCAP calibrated values fall below both the manual and raw ASPCAP results by 0.09 dex and 0.14 dex, respectively. The coolest RGB star in our sample has both raw and calibrated ASPCAP abundances that fall $\sim$0.15 dex below the manual value, with the manual abundance result agreeing with the abundances from the hotter giants: the manual O and Al abundances show no significant trend with T$_{\rm eff}$, while the ASPCAP results do. The abundances from Na, Si, and V exhibit similar behaviors among themselves, with the manual abundances falling in-between the calibrated and raw ASPCAP values. We note the large corrections to the raw ASPCAP abundances for Na, Si and V, becoming as large as $\sim$0.3 dex in the case of Na. Cobalt abundances from both raw and calibrated techniques seem to simply show larger scatter when compared to the manual analysis. The manually derived nitrogen abundances show marginal differences with the raw ASPCAP results, while the corrected ASPCAP abundances show good agreement. For titanium, the differences between the three sets of results are close to 0.1 dex with a similar abundance scatter.

As discussed previously, several spectroscopic investigations in the 1980's found that the metallicity of the open cluster NGC 2420 was around [Fe/H] = -0.6 dex (Pilachowski et al. 1980, [Fe/H] = -0.7 dex; Cohen 1980, [Fe/H ]= -0.6 dex; and Smith \& Suntzeff 1987, [Fe/H] = -0.5 dex). More recently, Pancino et al. (2010) analyzed several open clusters, including NGC 2420, using high-resolution (R = $\lambda$/$\delta$$\lambda$ $\approx$ 30,000) echelle optical spectra and found a metallicity for NGC 2420 that was near-solar, with [Fe/H] = -0.05 $\pm$ 0.03, therefore, much more metal-rich than the previous determinations. The study of Jacobson et al. (2011), using spectra obtained with the Hydra spectrograph on WIYN (R = $\lambda$/$\delta$$\lambda$ $\approx$ 18,000), found a metallicity of -0.20 $\pm$ 0.06. The mean iron abundance obtained here from the APOGEE spectra of 12 red-giants in NGC 2420 is $\langle$[Fe/H]$\rangle$ = -0.16 $\pm$ 0.04 and this result compares very well with the mean metallicity from Jacobson et al. (2011).

In addition, our analysis here has twelve chemical elements in common with Pancino et al. (2010) and Jacobson et al. (2011). Figure 7 provides a visual comparison of these results, shown as [X/Fe] versus [Fe/H].  Our abundances show small internal scatter in both [X/Fe] and [Fe/H], probably due to the high quality of the APOGEE spectra coupled to a homogeneous analysis.
Because Pancino et al. (2010) found a larger metallicity ([Fe/H]) than both this
study and Jacobson et al. (2011), all of the Pancino et al. points are shifted
to larger values of [Fe/H]; the Jacobson et al. (2011) iron abundances show larger scatter than ours, but generally overlap with our results.

Examining various element ratios ([X/Fe]) in Figure 7, the differences between the mean elemental abundances in the 3 studies are typically close to 0.1 dex, with a few points worth noting. Pancino et al. (2010) find two stars (from her sample of three)
that show somewhat higher values of
[O/Fe] and lower values of [Al/Fe]. There are offsets between the Jacobson et al. (2011) results and this study for almost all elements [O/Fe], [Mg/Fe], [Si/Fe], [Ca/Fe], and [Ti/Fe], except for sodium and nickel, which overlap almost perfectly. It is expected that these offsets are within the uncertainties from both stellar parameter determinations and gf-values.

Table 5 presents the final average chemical abundances from all stars
analyzed in NGC 2420 and their respective standard deviations.  The
derived standard deviations in all elements range from 0.02 - 0.05 dex,
well within expected uncertainties due to the abundance analysis itself.
The standard deviation values obtained limit any 
intrinsic abundance differences among this sample of red giants to less
than these rather small values: the observed red giants in NGC 2420 are
chemically homogeneous to a few hundredths of a dex.  Using a novel, and
very different technique, Bovy (2016) analyzed APOGEE spectra from 4
open clusters, including NGC 2420, to constrain abundance spreads in these
clusters. The technique removes T$_{\rm eff}$ trends in relative flux levels 
in both observed and simulated spectra and then evaluates the residuals both 
with, and without, abundance scatter in the simulated spectra. The
distributions of the values of the residuals can be used to provide
strong constraints on any underlying abundance variations in the cluster
stars.  Bovy (2016) finds quite small upper limits to any abundance
variations in all 4 clusters, including NGC 2420; values from Bovy (2016)
are included in Table 5. The upper limits set by Bovy (2016) compare well
with the limits set by the standard deviations resulting from the classical
spectroscopic abundance analysis performed here. The largest difference
between the two techniques for limiting abundance variations is for oxygen, from
OH, where here $\sigma$ = 0.03 dex, while the limit from Bovy (2016) is 0.06 dex.
The scatter found here is indeed small, given that OH is both sensitive to
T$_{\rm eff}$ and stellar metallicity (Table 4). Since the red
giants analyzed here have, except for one star, very similar temperatures
and the same metallicity, the small scatter found for oxygen may not be 
so surprising.

\subsection{Mixing in Red Giants}

The members of NGC 2420 present a useful combination
of stellar mass and metallicity for probing red giant
mixing along the RGB. With an estimated turn-off mass of
M $\sim$ 1.6M$_{\odot}$ and a metallicity of [Fe/H] = -0.16,
as measured here, the NGC 2420 red giants fall in a 
mass/metallicity range where the extent and impact of
non-standard mixing across the
luminosity bump is sensitive to the details of the type of
mixing and the input physics used in the modeling, 
(e.g. Charbonnel \& Lagarde 2010), Lagarde et al. 2012).
Of the elemental abundances analyzed here, it is $^{12}$C,
$^{14}$N, and the minor isotope $^{13}$C whose abundances
are most sensitive to both standard and non-standard mixing.
Eleven of the red giants in our study have
effective temperatures that are too hot (T$_{\rm eff}$ $\sim$ 4700 -- 4800 K) to 
easily measure the $^{13}$C$^{16}$O or
$^{13}$C$^{14}$N lines to strongly constrain values of
$^{12}$C/$^{13}$C, which is one of the most sensitive
indicators of extra-mixing.  The value of $^{12}$C/$^{14}$N,
however, can be used to probe extra-mixing, but is not as sensitive.  
Previous studies using APOGEE data (DR12) have used the [C/N] ratios
in order to estimate stellar masses and ages for the APOGEE sample
(Masseron \& Gilmore 2015; Ness et al. 2016 and Martig et al. 2016).

Assuming initial scaled-solar values of [C/Fe] 
and [N/Fe] for NGC 2420 (since it is only slightly sub-solar
in metallicity, this assumption is a likely good approximation),
the red giants measured here have slightly lowered mean values
of [$^{12}$C/Fe] = -0.06 and elevated values of [$^{14}$N/Fe] = +0.11, which are what is expected qualitatively for first dredge-up in low-mass red giants.
The altered $^{12}$C and $^{14}$N abundances are due to H-burning
on the CN-cycle, as predicted by stellar evolution,
with the result that the total number of CNO nuclei are
conserved.  Neglecting $^{13}$C, which is a minor isotope, the
approximate conservation of $^{12}$C + $^{14}$N nuclei can be tested
in these red giants, under the assumption that initial abundance ratios
were [C/Fe] = 0.0 and [N/Fe] = 0.0. The NGC 2420 red giants are identified in Figure 8
as either RGB or RC stars (see discussion in Section 3), with the
error bars equal to the standard deviations of the means from each
abundance determination.  The hotter red giants, near the lower RGB 
and RC, scatter around the C+N curve quite closely: within less than 0.1 dex, which is
similar to the expected uncertainties.  These red giants display the
signature of the first dredge-up of matter exposed to the CN-cycle.
The coolest red giant analyzed here, 2M07381507+2134589, is offset
from the hotter giants, as well as the C+N curve.  This offset ($\sim$0.1) is
relatively small by typical abundance standards; however, given the accuracy of the analysis of APOGEE spectra, it is significantly larger than the abundance uncertainties.  This effect for carbon abundances, as derived from CO molecular lines, has been noted in NGC 6791 from APOGEE spectra (Cunha et al. 2015), with the result that carbon abundances decrease by $\sim$0.1 dex from T$_{\rm eff}$ $\sim$ 5000 K to 4000 K. For the discussion here, this red giant is not considered in constraining stellar models from its $^{12}$C abundance alone.

In Figure 8 the two groups of red giants (RGB and RC) do not show obvious
differences in their respective C and N abundances. The mean abundances are 
$\langle$A($^{12}$C$_{\rm RGB}$)$\rangle$ = 8.17 $\pm$ 0.03 and 
$\langle$A($^{14}$N$_{\rm RGB}$)$\rangle$ = 7.77 $\pm$ 0.03 for the five RGB stars (we
do not include the coolest RGB star) and the mean 
values of six RC stars are A($^{12}$C$_{\rm RC}$) = 8.18 $\pm$ 0.02 and 
A($^{14}$N$_{\rm RC}$) = 7.80 $\pm$ 0.04.  The corresponding mean
values of $^{12}$C/$^{14}$N for the RGB and RC stars are, respectively,
2.50 $\pm$ 0.29 and 2.36 $\pm$ 0.18.  We note that the RC mean value 
of C/N is slightly smaller than for the lower RGB stars, which would be 
in the sense of extra-mixing. However, this difference is not statistically significant or conclusive. We note, however, that differences for C/N between RC and RGB stars have also been reported by Mikolaitis et al. (2012) and Drazdauskas et al. (2016), who obtain ratios of  (C/N$_{\rm RC}$ = 1.62, C/N$_{\rm RGB}$ = 2.04) and (C/N$_{\rm RC}$ = 1.60, C/N$_{\rm RGB}$ = 1.74)  for the open cluster Collinder 261. In addition, Tautvai{\v s}iene et al. (2000) obtained C/N$_{\rm RC}$ = 1.40 and C/N$_{\rm RGB}$ = 1.70 for M67. These three studies all find somewhat lower C/N ratios on the RC when compared with the RGB. On a more quantitative footing, the results here constrain any extra-mixing, between the lower RGB through the He core-flash and onto the RC, causing a $\Delta$(C/N) to be less than 0.1-0.3 in the linear ratio.

Recent studies using the previous APOGEE data release (DR12) have used the [C/N] ratios in order to estimate stellar masses and ages for the APOGEE sample
(Masseron \& Gilmore 2015; Ness et al. 2016; Martig et al. 2016). The results from Martig et al. (2016), would indicate a mean mass for our NGC 2420 sample of M$_{\star}$ $\sim$ 1.31 $\pm$ 0.12 M$_{\odot}$ and a mean age of $\sim$ 3.56 $\pm$ 0.86 Gyr, therefore, finding this open cluster to be older than what we adopt. The mean masses and ages for the studied stars estimated in Ness et al. (2016) are: M$_{\star}$ = 1.52 $\pm$ 0.22 M$_{\odot}$ and age $\sim$ 2.84 $\pm$ 0.86 Gyr. However, both these studies are based on DR12 and the improved abundances from DR13 have not yet been adopted.

\subsection{Abundance Comparisons with Galactic Trends}

Results for the Milky Way field disk stars, defining the Galactic trends, are also shown as comparisons in Figure 9. We use the results from Adibekyan et al. (2012; blue circles), Bensby et al. (2014; green triangles), Allende Prieto et al. (2004; magenta squares), Nissen et al. (2014; cyan pentagons), Reddy et al. (2003; grey axis) and Carretta et al (2000; black pluses), to define the disk trends. The abundances obtained for our sample of red giants in NGC 2420 are in general agreement with what is obtained for field disk stars at the corresponding metallicity of NGC 2420, although the derived abundances of, for example, Mg, Ca, Ti, V, and Co, show some marginal systematic differences when compared to field star results shown in Figure 9; these fall close to the lower envelope of the elemental distribution obtained in the other studies. Some of those samples are quite local to the solar neighborhood, such as, Allende Prieto et al. (2004) who have stars within a volume within 15 pc from the Sun, while other samples extend much further into the disk, as well as the thick disk (Bensby et al. 2014). In addition, there is a metallicity gradient in the Milky Way disk. Several recent studies derive metallicity gradients from open clusters (Cunha et al. 2016, Frinchaboy et al. 2013, Jacobson et al. 2011, Andreuzzi et al. 2011, Carrera $\&$ Pancino 2011, Magrini et al. 2009). For APOGEE results, in particular, Cunha et al. (2016) present metallicity gradients based on DR12 abundances of 29 open clusters. The obtained gradients of [X/H] are typically -0.030 dex/kpc with some possible evidence of flatter gradients for R$_{\rm GC}$ $>$ 12 kpc. Having a Galactocentric distance R$_{\rm GC}$ $\sim$ 11 kpc, the derived abundances here are in line, i.e., about 0.1 dex lower, with the derived gradients from the APOGEE open cluster sample results in DR12, although for some elements there are small systematic offsets due to the different line lists used here and in DR12.

\section{Summary}

A manual abundance analysis was carried out for the open cluster NGC 2420 using APOGEE spectra. Twelve red giants (6 from the RGB and 6 from the RC) were included with abundances derived for 16 chemical elements.  A comparison between the manually derived stellar parameters and abundances with those from ASPCAP found overall good agreement in T$_{\rm eff}$ and log g, when the ASPCAP calibrated surface gravities were used.  Good agreements (i.e., $\le$0.1 dex) in the chemical abundances were found for many elements, although some exhibit larger offsets between ASPCAP raw or calibrated abundances when compared to manual values; the most notable differences are for Na, Al, V, and to a lesser extent Si.

The mean iron abundance and standard deviation of the mean were found to be [Fe/H] = -0.16 $\pm$ 0.04, which is in good agreement with the recent result from Jacobson et al. (2011) of [Fe/H] = -0.20 based on optical spectra. This value for [Fe/H] is in-line with what would be expected for a cluster 2 kpc farther out from the solar circle given an Fe-abundance gradient of $\sim$ -0.03 dex/kpc (Cunha et al. 2016).  Values of [X/Fe] for the other elements do not deviate significantly from solar (not including C and N which are affected by the first dredge-up), although [O/Fe], [Na/Fe], and [Al/Fe] are slightly elevated by $\sim$+0.1 dex, while [Co/Fe] $\sim$ -0.1. We note that the values for [O/Fe] and [Al/Fe] follow the trends defined by the Galactic thin disk stars, while, [Na/Fe] remains somewhat offset from the Galactic trend possibly due to systematic differences in the studies.

The NGC 2420 red giants have $^{12}$C and $^{14}$N abundances that are consistent with first dredge-up, with carbon-12 mildly depleted (-0.06 dex on average) and nitrogen-14 slightly elevated (+0.11 dex on average).  No significant differences in the ratio of $^{12}$C/$^{14}$N are found between the RGB stars (below the luminosity bump) and the RC stars, providing some constraints on extra-mixing mechanisms.  More stringent tests of possible extra-mixing will come with an analysis of $^{12}$C/$^{13}$C, which is difficult in the APOGEE spectral window for red giants having effective temperatures that are characteristic of the lower RGB and RC in NGC 2420.

\section{Acknowledgments}

We thank H-W. Rix and A. Roman-Lopes for useful comments. Funding for the Sloan Digital Sky Survey IV has been provided by the Alfred P. Sloan Foundation, the U.S. Department of Energy Office of
Science, and the Participating Institutions. SDSS-IV acknowledges
support and resources from the Center for High-Performance Computing at
the University of Utah. The SDSS web site is www.sdss.org.

SDSS-IV is managed by the Astrophysical Research Consortium for the 
Participating Institutions of the SDSS Collaboration including the 
Brazilian Participation Group, the Carnegie Institution for Science, 
Carnegie Mellon University, the Chilean Participation Group, the French Participation Group, Harvard-Smithsonian Center for Astrophysics, 
Instituto de Astrof\'isica de Canarias, The Johns Hopkins University, 
Kavli Institute for the Physics and Mathematics of the Universe (IPMU) / 
University of Tokyo, Lawrence Berkeley National Laboratory, 
Leibniz Institut f\"ur Astrophysik Potsdam (AIP),  
Max-Planck-Institut f\"ur Astronomie (MPIA Heidelberg), 
Max-Planck-Institut f\"ur Astrophysik (MPA Garching), 
Max-Planck-Institut f\"ur Extraterrestrische Physik (MPE), 
National Astronomical Observatory of China, New Mexico State University, 
New York University, University of Notre Dame, 
Observat\'orio Nacional / MCTI, The Ohio State University, 
Pennsylvania State University, Shanghai Astronomical Observatory, 
United Kingdom Participation Group,
Universidad Nacional Aut\'onoma de M\'exico, University of Arizona, 
University of Colorado Boulder, University of Oxford, University of Portsmouth, 
University of Utah, University of Virginia, University of Washington, University of Wisconsin, 
Vanderbilt University, and Yale University. CAP acknowledges support from the Spanish
MINECO through grant AYA2014-56359-P. D.A.G.H. was funded by the Ramóon y Cajal fellowship number  RYC-2013-14182. D.A.G.H. and O.Z. acknowledge support  provided  by  the  Spanish  Ministry  of  Economy  and  Competitiveness (MINECO) under grant AYA-2014-58082-P. RC acknowledge support from the Spanish Ministry of Economy and Competitiveness through grants AYA2014-56359-P, AYA2014-56795, and AYA2013-42781P. Szabolcs M\'esz\'aros has been supported by the J\'anos Bolyai Research Scholarship of the Hungarian Academy of Sciences. B.T., S.V. and D.G. gratefully acknowledge support from the Chilean BASAL Centro de Excelencia en Astrof\'isica y Tecnolog\'ias Afines (CATA) grant PFB-06/2007. P.M.F. acknowledges support from NSF Grant AST-1311835. ARL thanks partial support from the DIULS Regular Project PR15143. 


\appendix

\clearpage

\newpage
\begin{figure}
\epsscale{0.8}
\plotone{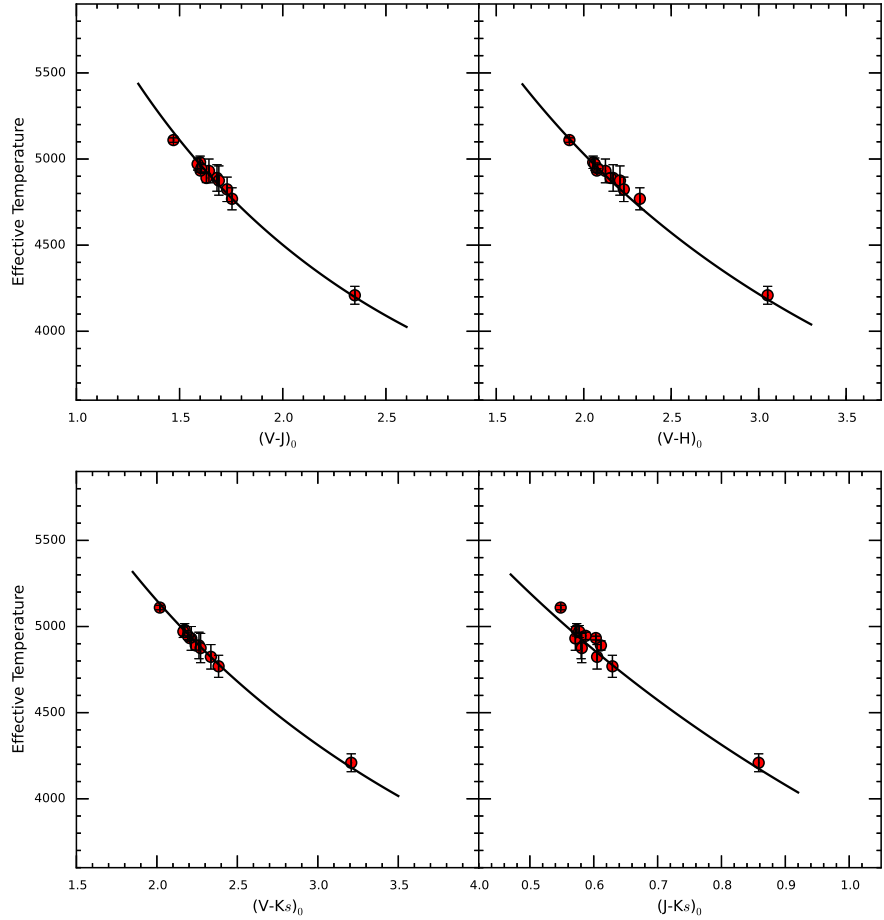}
\caption{The effective temperature calibrations from Gonz\'alez-Hern\'andez $\&$ Bonifacio (2009) for the different colors considered in this study: ($V$-$J$)$_{0}$, ($V$-$H$)$_{0}$, ($V$-$K$$_{\rm S}$)$_{0}$, and ($J$-$K$$_{\rm S}$)$_{0}$. The red points are our derived effective temperatures.}
\end{figure}

\newpage
\begin{figure}
\epsscale{0.7}
\plotone{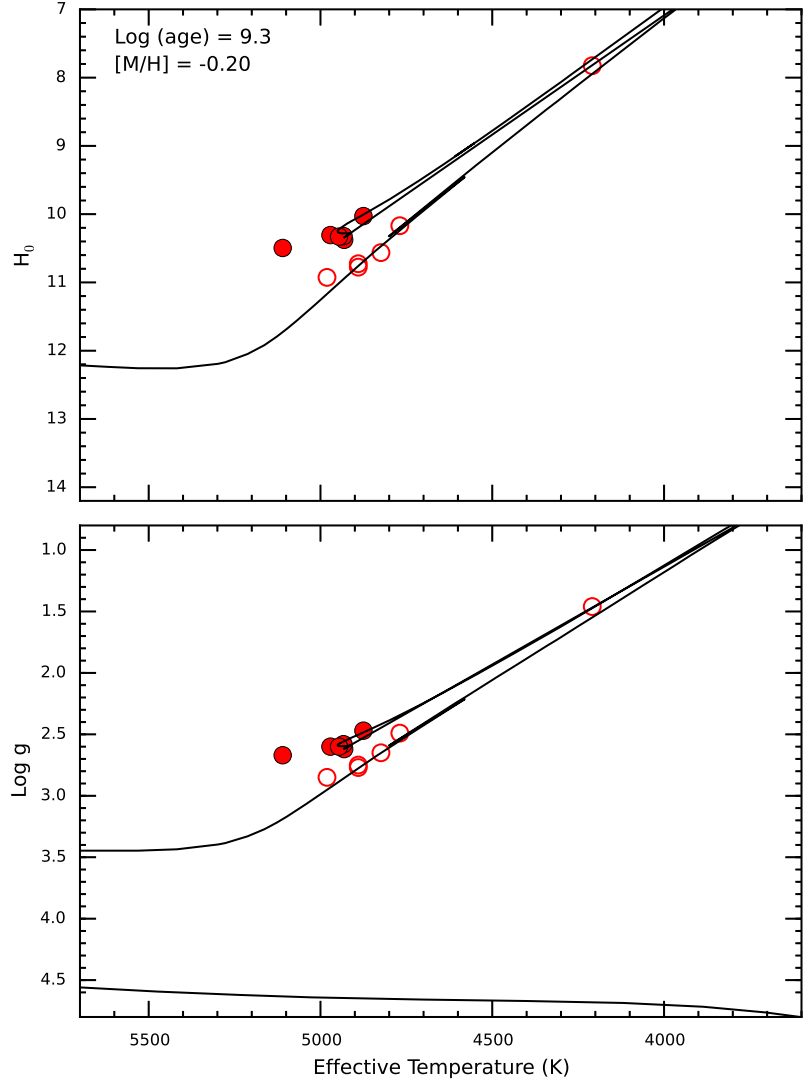}
\caption{Top panel: T$_{\rm eff}$ vs H$_{0}$ magnitude diagram. Black line indicates a 2 Gyr isochrone from Bressan et al. (2012) for [M/H] = -0.20 dex. Our sample contains six stars on the red-giant branch (open circles) and 6 are red-clump giants (filled circles). Bottom panel: HR-diagram with the surface gravities obtained from the fundamental relation in Equation 1.}
\end{figure}

\newpage
\begin{figure}
\epsscale{1}
\plotone{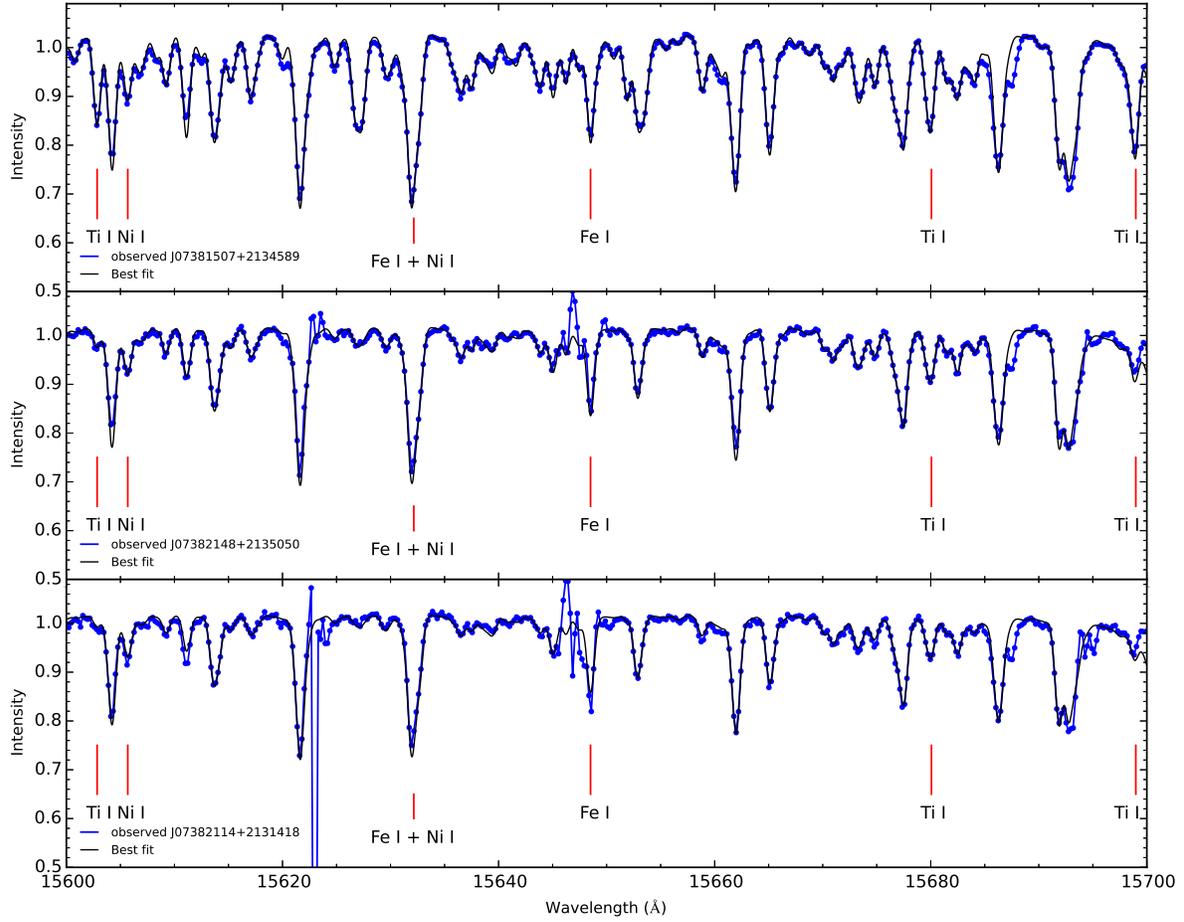}
\caption{Sample observed spectra for three stars along with the best fit synthetic spectra. Lines of Ti I, Fe I and Ni I used in the abundance analysis are indicated. The star with the lowest effective temperature  (J07381507+2134589; T$_{\rm eff}$ = 4209 K) is shown in the top panel. The middle panel shows a red-clump star (J07382148+2135050; T$_{\rm eff}$ = 4890 K). The hottest star in our sample (J07382114+2131418; T$_{\rm eff}$ = 5111 K) is shown the bottom panel.}
\end{figure}

\newpage
\begin{figure}
\epsscale{0.8}
\plotone{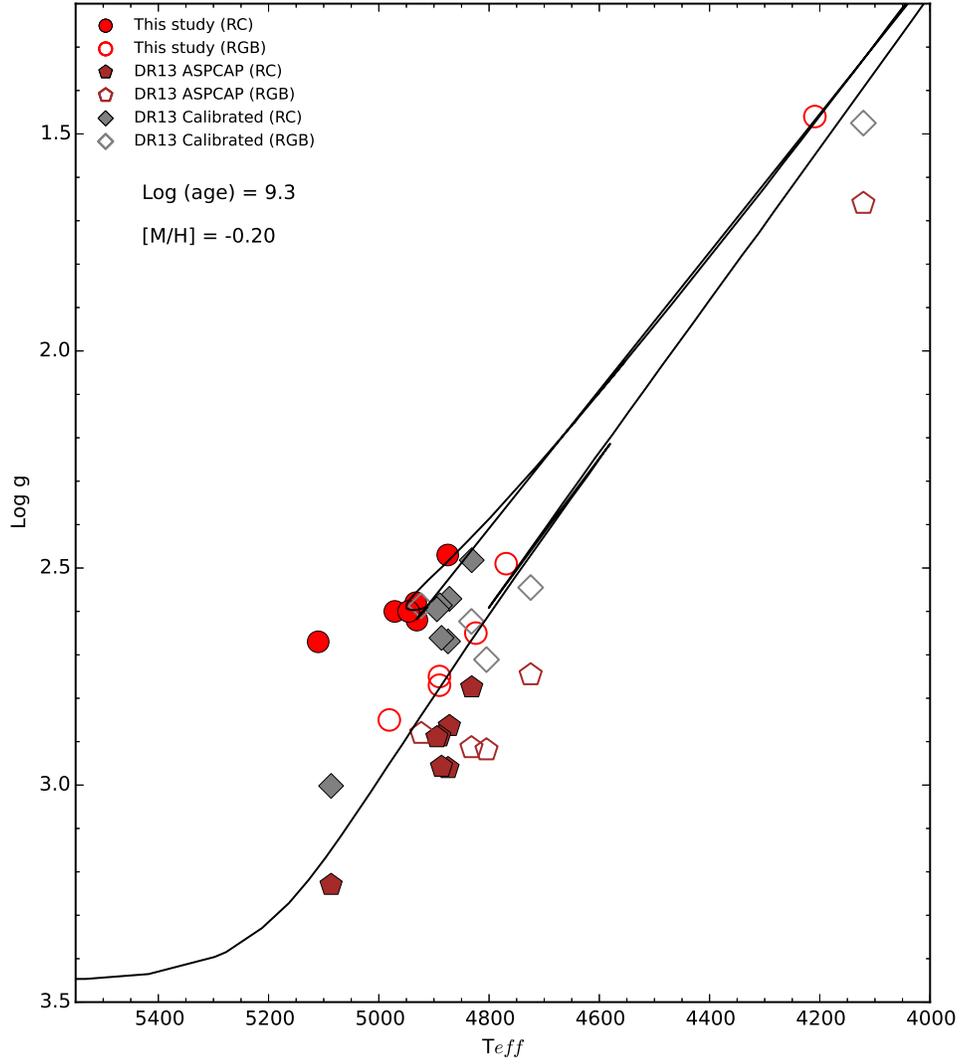}
\caption{HR-diagram of log g versus T$_{\rm eff}$ and showing a PARSEC isochrone from Bressan et al. (2012). Results from this study are compared with those from ASPCAP both for raw and calibrated values.}
\end{figure}

\newpage
\begin{figure}
\epsscale{0.8}
\plotone{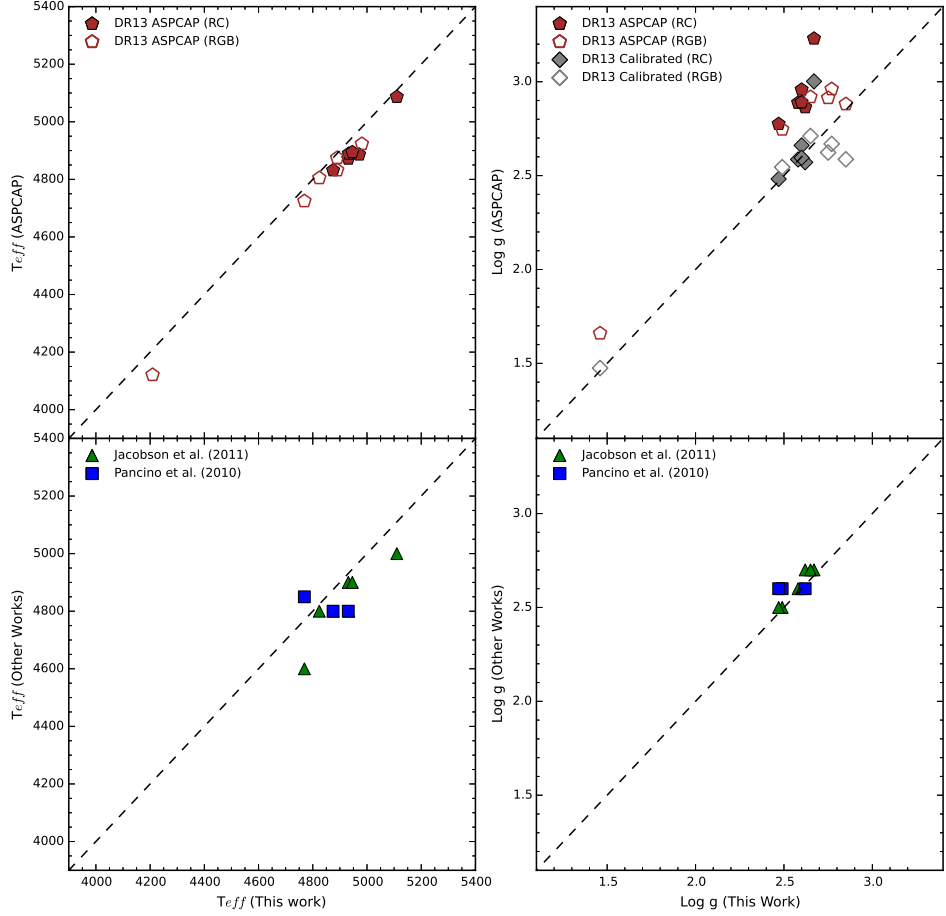}
\caption{Comparisons of the atmospheric parameters derived here with the results obtained with the APOGEE abundance pipeline ASPCAP (top panels) and the literature (bottom panels).  The open symbols in the top panels represent stars on the RGB, while the filled symbols in the top panels are red clump stars. The left panels show the comparisons for the effective temperatures, while the right panels show surface gravity comparisons. Note that the log g's derived from Equation 1 agree very well with the calibrated ASPCAP values.}
\end{figure}

\newpage
\begin{figure}
\epsscale{1}
\plotone{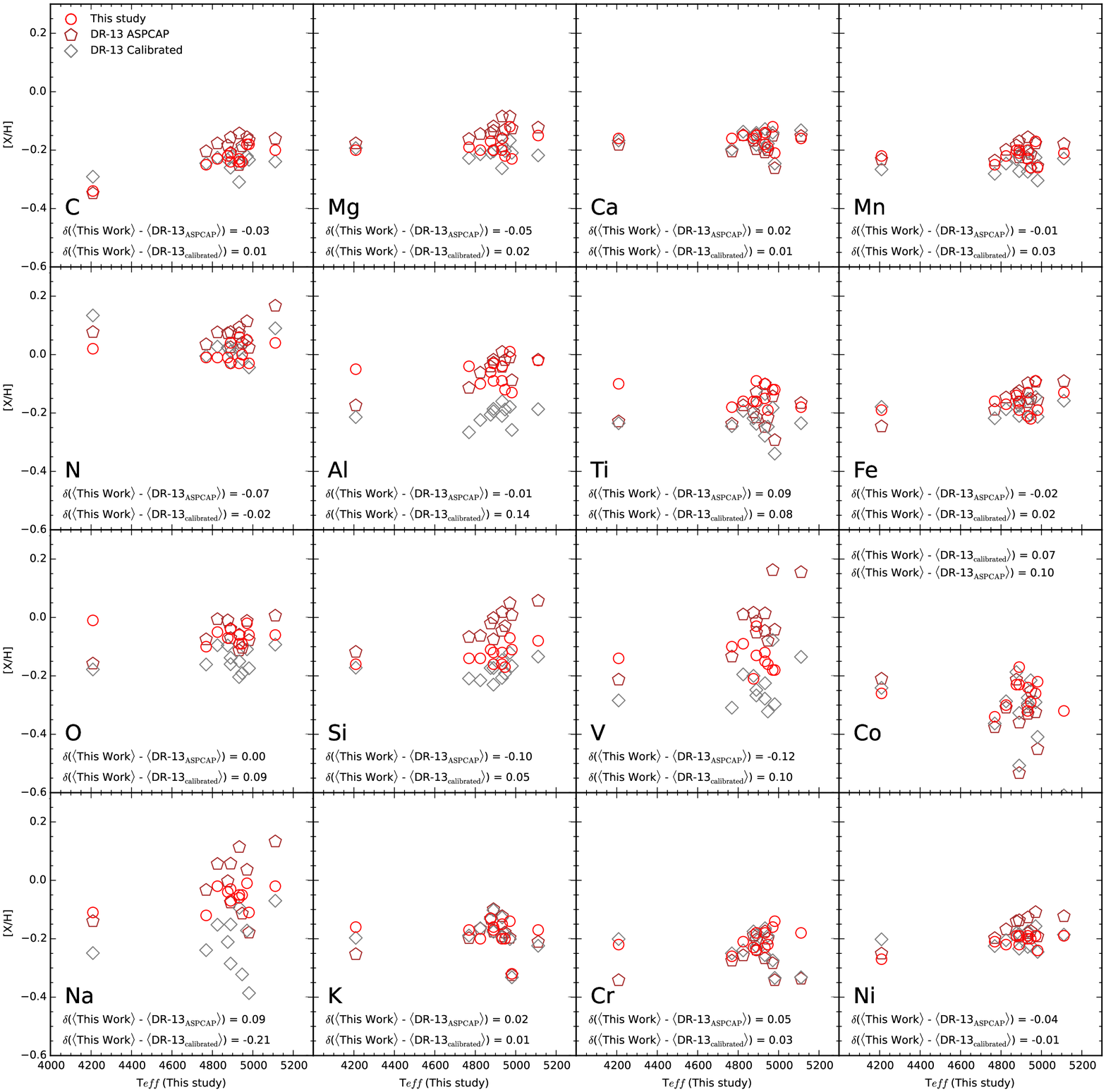}
\caption{Chemical abundances for all elements are shown as a function of T$_{\rm eff}$ for three sets of results: abundances from the manual analysis in this study; raw ASPCAP abundances and calibrated ASPCAP DR13 abundances.}
\end{figure}

\newpage
\begin{figure}
\epsscale{1}
\plotone{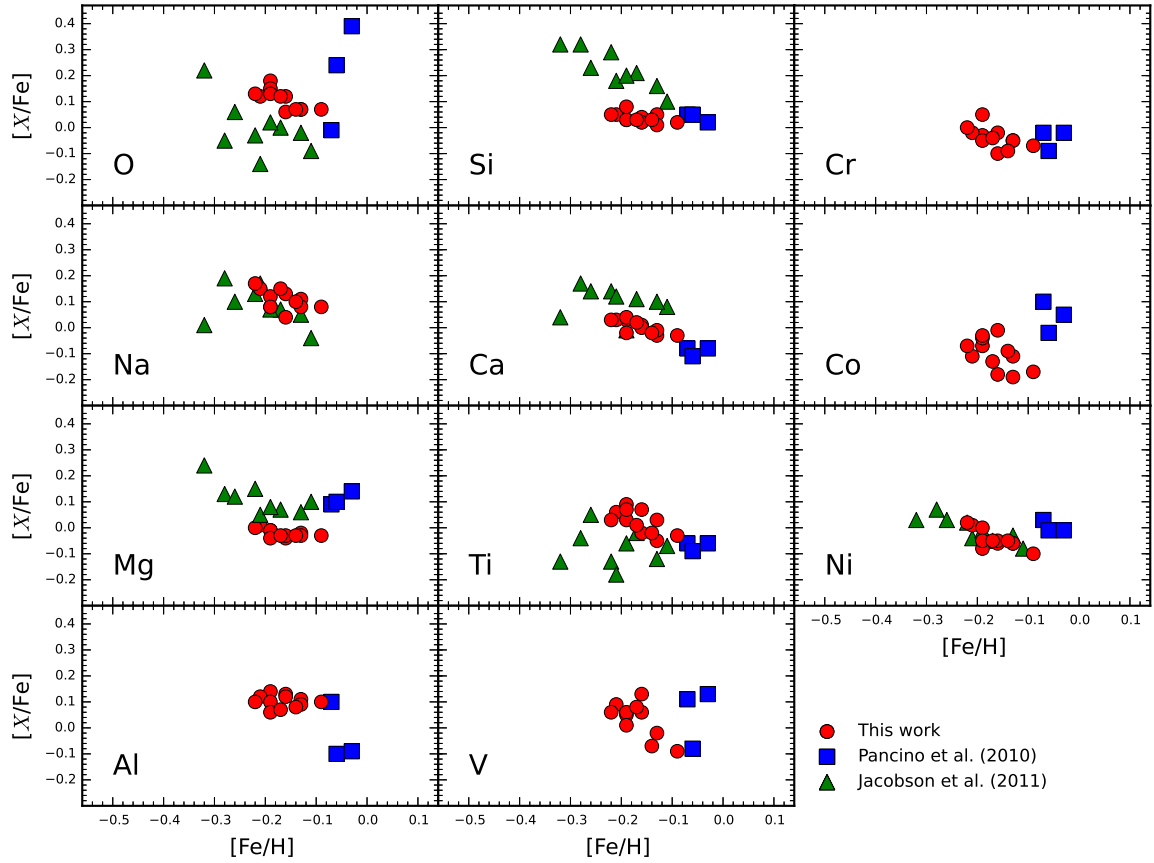}
\caption{Values of  [X/Fe] versus [Fe/H] for elements analyzed in common with the studies of Pancino et al. (2010) and Jacobson et al. (2011).}
\end{figure}

\newpage
\begin{figure}
\epsscale{0.8}
\plotone{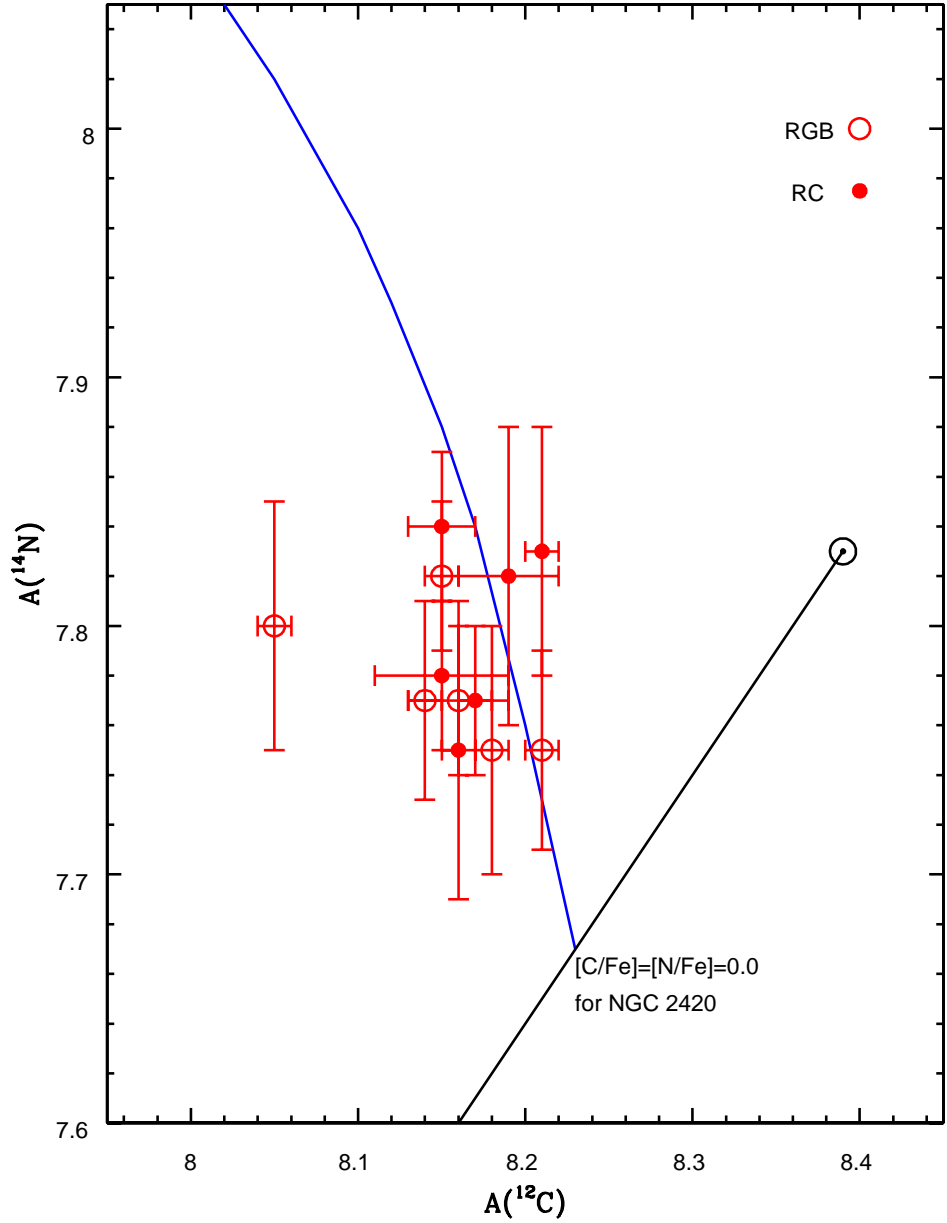}
\caption{A($^{14}$N) versus A($^{12}$C) for the six RGB and six RC sample stars showing the signature of first dredge-up. The solar symbol indicates the solar abundances adopted by APOGEE/ASPCAP (Asplund et al. 2005), with the solid black line delineating scaled-solar values of A($^{12}$C) and A($^{14}$N). The solid blue curve represents a constant sum of $^{12}$C and $^{14}$N, as carbon is cycled into nitrogen, beginning with initial values scaled down from the Sun by -0.16 dex.}
\end{figure}

\newpage
\begin{figure}
\epsscale{1}
\plotone{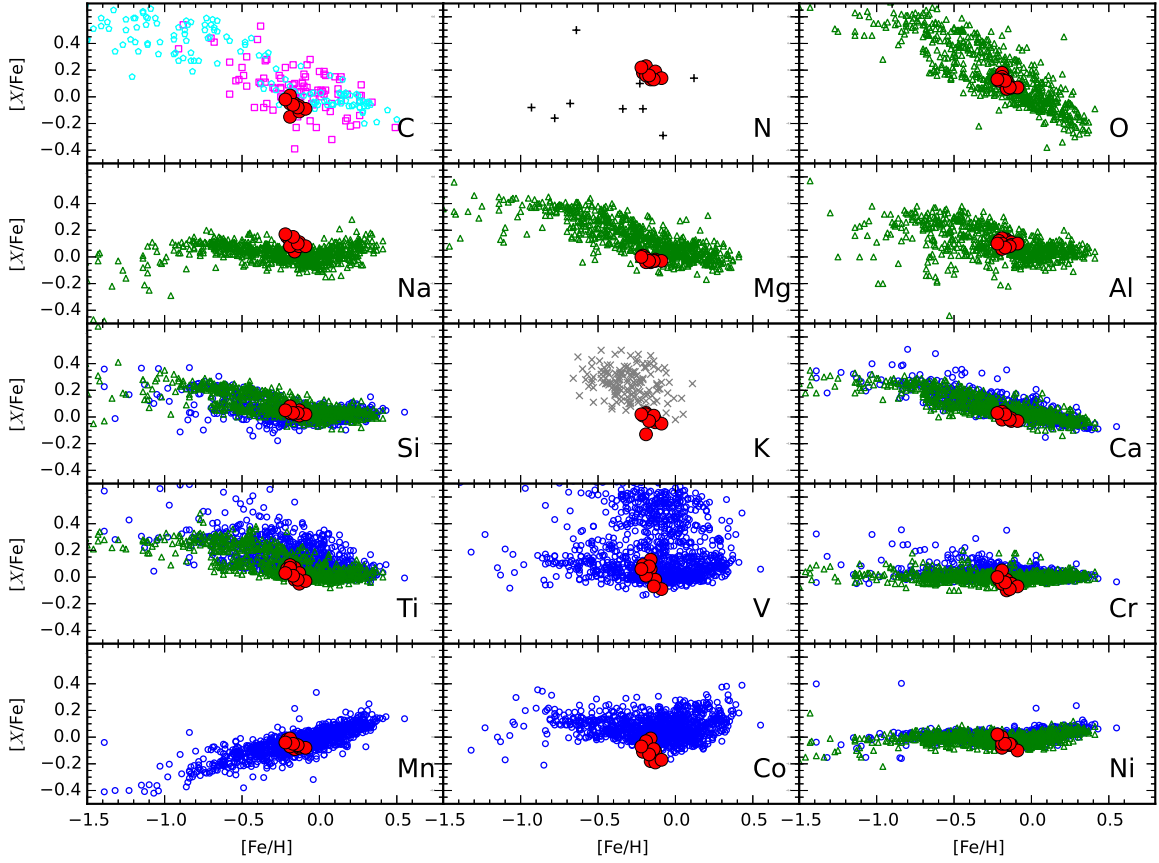}
\caption{Galactic trends of [X/Fe] as a function of [Fe/H] for the stars in the open cluster NGC 2420 (red points). Field stars in the thin and thick disk are from Bensby et al. (2014, green triangle); Adibekyan et al. (2012, blue circles), Allende-Prieto et al. (2004, magenta squares), Nissen et al. (2014; cyan pentagons), Reddy et al. (2003; grey axis) and Carreta et al (2000; black pluses).}
\end{figure}

\clearpage
\begin{deluxetable}{lccccccccccc}
\tabletypesize{\scriptsize}
\rotate
\tablecaption{Atmospheric Parameters}
\tablewidth{0pt}
\tablehead{
\colhead{Star} & \colhead{RV} & \colhead{SNR} & \colhead{$V$} & \colhead{$J$} & \colhead{$H$} & \colhead{$K$} &  \colhead{T$_{\rm eff}$ (K)} & \colhead{log g (cm-s$^{-2}$)} & \colhead{$\xi$ (km-s$^{-1}$)}  & \colhead{Note} 
}

\startdata

J07380545+2136507 							&	73.49 $\pm$ 0.56	& 220  &	13.06  &11.258	&10.755	&10.651	&4890 $\pm$ 77   &	2.75 &	1.40	&	RGB	\\
J07380627+2136542 $^{\dagger}$	$^{\star}$	&	73.93 $\pm$ 0.56	& 414  &	12.656 &10.781	&10.198	&10.125	&4769 $\pm$ 64   &	2.49 &	1.45	&	RGB	\\
J07381507+2134589							&	74.39 $\pm$ 0.29	& 1089 &	11.042 &8.572	&7.854	&7.687	&4209 $\pm$ 53   &	1.46 &	1.60	&	RGB	\\
J07381549+2138015 $^{\dagger}$	$^{\star}$	&	74.59 $\pm$ 0.48	& 323  &	12.666 &10.903	&10.405	&10.305	&4932 $\pm$ 69	 &	2.62 &	1.80	&	RC	\\
J07382114+2131418 $^{\dagger}$				&	74.24 $\pm$ 0.28	& 138  &	12.579 &10.988	&10.524	&10.413	&5111 $\pm$ 14	 &	2.67 &	1.60	&	RC	\\
J07382148+2135050							&	74.13 $\pm$ 1.18	& 222  &	13.096 &11.345	&10.805	&10.707	&4890 $\pm$ 27   &	2.77 & 	1.80	&	RGB	\\
J07382195+2135508 $^{\dagger}$				&	73.58 $\pm$ 0.14	& 272  & 	12.562 &10.840	&10.350	&10.210	&4933 $\pm$ 11   &	2.58 &	1.70	&	RC	\\
J07382347+2124448							&	74.14 $\pm$ 0.38	& 117  & 	13.133 &11.426	&10.955	&10.826	&4981 $\pm$ 36   &	2.85 &	1.70	&	RGB	\\
J07382670+2128514							&	74.38 $\pm$ 0.33	& 131  &	12.535 &10.827	&10.335	&10.223	&4971 $\pm$ 35   &	2.60 &	1.50	&	RC	\\
J07382696+2138244 $^{\dagger}$	$^{\star}$	&	73.67 $\pm$ 1.12	& 316  &	12.401 &10.590	&10.057	&9.982	&4876 $\pm$ 85	 &	2.47 &	1.60	&	RC	\\
J07382984+2134509 $^{\dagger}$				&	75.11 $\pm$ 0.22	& 321  &	12.958 &11.107	&10.592	&10.475	&4825 $\pm$ 71	 &	2.65 &	1.40	&	RGB	\\
J07383760+2134119 $^{\dagger}$				&	73.83 $\pm$ 0.23	& 270  &	12.574 &10.848	&10.358	&10.234	&4947 $\pm$ 31	 &	2.60 &	2.00	&	RC	\\

\enddata
\tablenotetext{1}{ $\dagger$ Stars in common with Jacobson et al. (2011)}
\tablenotetext{2}{ $\star$ Stars in common with Pancino et al. (2010)}
\end{deluxetable}

\begin{deluxetable}{cccccccccccccc}
\rotate
\tabletypesize{\tiny}
\tablecaption{Molecular Lines and Derived Abundances}
\tablewidth{0pt}
\tablehead{
\colhead{Element} & \colhead{ $\lambda$ (\AA{})}  &  \colhead{ J07380545} & \colhead{ J07380627} & \colhead{ J07381507} & \colhead{ J07381549} & \colhead{ J07382114} & \colhead{ J07382148} & \colhead{ J07382195} & \colhead{ J07382347} & \colhead{J07382670} & \colhead{ J07382696} & \colhead{J07382984} & \colhead{J07383760} \\ 
\colhead{} & \colhead{} & \colhead{ +2136507} & \colhead{ +2136542} & \colhead{ +2134589} & \colhead{ +2138015} & \colhead{+2131418} & \colhead{ +2135050} & \colhead{ +2135508} & \colhead{ +2124448} & \colhead{+2128514} & \colhead{ +2138244} & \colhead{ +2134509} & \colhead{+2134119} }
\startdata

 {\bf $^{12}$C from $^{12}$C$^{16}$O} & & & & & & & & &\\
 
 (4-1) V-R       & 15774. &8.19	 &8.12	 &8.06	 &8.15	 &8.19	 &8.15	 &8.14	 &8.20	 &8.21	 &8.16	 &8.16	 &8.17\\
 (5-2) V-R       & 15976. &8.17	 &8.15	 &8.04	 &8.17	 &8.16	 &8.14	 &8.13	 &8.23	 &8.21	 &8.16	 &8.13	 &8.18\\
 (6-3) V-R       & 16183. &8.17	 &8.15	 &8.05	 &8.17	 &8.23	 &8.15	 &8.18	 &8.20	 &8.22	 &8.18	 &8.20	 &8.10\\
 $\langle$A(C)$\rangle$ $\pm$ $\sigma$ & & 8.18 $\pm$ 0.01 &	8.14 $\pm$ 0.01	&8.05 $\pm$ 0.01 &	8.16 $\pm$ 0.01	& 8.19 $\pm$ 0.03 &	8.15 $\pm$ 0.01	& 8.15 $\pm$ 0.02 & 	8.21 $\pm$ 0.01	& 8.21 $\pm$ 0.01	& 8.17 $\pm$ 0.01	& 8.16 $\pm$ 0.03 & 8.15 $\pm$ 	0.04\\
 & & & & & & & & & & & &\\

{\bf $^{16}$O from $^{16}$OH} & & & & & & & & & \\
 (2-0) P$_{1}$ 9.5    & 15278.   &...	 &...	 &8.70	 &...	 &...	 &...	 &...	 &...	 &...	 &...	 &...	 &...\\
 (2-0) P$_{1}$ 9.5    & 15281. 	&8.65	 &8.53	 &8.64	 &8.50	 &8.58	 &8.69	 &8.61	 &8.58	 &8.68	 &8.59	 &8.64	 &8.57\\
 (3-1) P$_{2}$ 3.5    & 15390.  	&8.57	 &8.56	 &8.62	 &8.62	 &...	 &8.62 	 &...	 &8.62	 &8.63	 &8.56	 &8.61	 &8.61\\
 (2-0) P$_{2}$ 11.5   & 15568.  	&8.61	 &8.63	 &8.65	 &8.62	 &8.63	 &8.58	 &8.65	 &8.59	 &8.63	 &8.62	 &8.63	 &8.56\\                                          
 (3-1) P$_{2}$ 9.5    & 16190.  	&8.66	 &8.56	 &8.63	 &8.56	 &8.58	 &8.61	 &8.56	 &8.59	 &8.61	 &8.54	 &8.63	 &8.55\\
 (3-1) P$_{2}$ 9.5    & 16192.  	&8.62	 &8.55	 &8.63	 &8.57	 &...	 &8.61	 &8.59	 &...	 &8.66	 &8.62	 &8.54	 &...\\
$\langle$A(O)$\rangle$ $\pm$ $\sigma$ & & 8.62 $\pm$ 0.03	 &  8.56 $\pm$ 0.04	 &8.65 $\pm$ 0.03 &	8.57 $\pm$ 0.04 &	8.60 $\pm$ 0.02 &	8.62 $\pm$ 0.05 &	8.60 $\pm$ 0.03 &	8.60 $\pm$ 0.02	 &8.64 $\pm$ 0.02 &8.59 $\pm$ 0.03	 &8.61 $\pm$ 0.03	 & 8.57 $\pm$ 0.02\\
 & & & & & & & & & & & &\\                                                                                       
{\bf $^{14}$N from $^{12}$C$^{14}$N} & & & & & & & & & \\
 (1-2) Q2 41.5         & 15260.	&...	 &7.74	 &7.69	 &7.66	 &7.73	 &7.77	 &7.80	 &7.75	 &7.73	 &7.72	 &7.72	 &7.75\\
 (1-2) P2 34.5         & 15322.	&7.79	 &7.77	 &7.77	 &7.71	 &7.86	 &7.85	 &7.83	 &7.75	 &7.87	 &7.78	 &7.76	 &7.78\\
 (1-2) R2 56.5         & 15397.	&7.76	 &7.80	 &7.85	 &7.81	 &7.89	 &7.85	 &7.86	 &7.77	 &7.84	 &7.83	 &7.76	 &7.78\\
 (0-1) R1 68.5         & 15332.	&...	 &7.77	 &7.83	 &7.83	 &7.80	 &7.83	 &7.88	 &...	 &7.88	 &7.79	 &7.78	 &7.85\\
 (0-1) P2 49.5         & 15410.	&7.75	 &7.80	 &7.78	 &7.79	 &7.86	 &7.79	 &7.85	 &7.70	 &7.84	 &7.75	 &7.76	 &7.77\\	
 (0-1) Q2 59.5         & 15447.	&7.82	 &7.82	 &7.87	 &7.77	 &7.88	 &7.89	 &7.84	 &7.82	 &7.89	 &7.77	 &7.81	 &7.81\\
 (0-1) Q1 60.5         & 15466.	&7.76	 &7.81	 &7.82	 &7.74	 &7.79	 &7.79	 &7.79	 &7.78	 &7.83	 &7.78	 &7.79	 &7.72\\
 (1-2) P2 38.5         & 15472.	&7.71	 &7.69	 &7.80	 &7.65	 &...	 &7.85	 &7.85	 &7.66	 &7.78	 &7.71	 &7.72	 &7.77\\
 (0-1) P1 51.5         & 15482.	&7.66	 &7.76	 &7.79	 &7.78	 &7.71	 &7.83	 &7.89	 &7.74	 &7.84	 &7.76	 &7.83	 &7.75\\
$\langle$A(N)$\rangle$ $\pm$ $\sigma$ & & 7.75 $\pm$ 0.05 &7.77 $\pm$ 0.04&	7.80 $\pm$ 0.05&	7.75 $\pm$ 0.06&	7.82 $\pm$ 0.06&	7.82 $\pm$ 0.03&	7.84 $\pm$ 0.03&	7.75 $\pm$ 0.04&	7.83 $\pm$ 0.05&	7.77 $\pm$ 0.03&	7.77 $\pm$ 0.03&	7.78 $\pm$ 0.03\\
 & & & & & & & & & & & &\\

\enddata
\end{deluxetable}

\clearpage

\begin{deluxetable}{cccccccccccccc}
\rotate
\tabletypesize{\tiny}
\tablecaption{Atomic Lines and Derived Abundances}
\tablewidth{0pt}
\tablehead{
\colhead{Element} & \colhead{$\lambda$ (\AA{})}  &  \colhead{ J07380545} & \colhead{ J07380627} & \colhead{ J07381507} & \colhead{ J07381549} & \colhead{ J07382114} & \colhead{ J07382148} & \colhead{ J07382195} & \colhead{ J07382347} & \colhead{J07382670} & \colhead{ J07382696} & \colhead{J07382984} & \colhead{J07383760} \\ 
\colhead{} & \colhead{} & \colhead{ +2136507} & \colhead{ +2136542} & \colhead{ +2134589} & \colhead{ +2138015} & \colhead{+2131418} & \colhead{ +2135050} & \colhead{ +2135508} & \colhead{ +2124448} & \colhead{+2128514} & \colhead{ +2138244} & \colhead{ +2134509} & \colhead{+2134119} }

\startdata

{\bf Fe I}  	  & 15194.492	 &7.40	 &7.33	 &7.26	 &7.32	 &7.38	 &7.34	 &7.42	 &7.35	 &7.48	 &7.38	 &7.35	 &7.32\\
				  & 15207.526	 &7.24	 &7.37	 &7.23	 &7.16	 &7.32	 &7.17	 &7.27	 &7.13	 &7.34	 &7.36	 &7.29	 &7.19\\
				  & 15395.718	 &7.24	 &7.26	 &7.24	 &7.23	 &7.29	 &7.25	 &7.34	 &7.20	 &7.36	 &7.29	 &7.27	 &7.19\\
				  & 15490.339	 &7.37	 &7.25	 &7.23	 &7.30	 &7.26	 &7.33	 &7.33	 &7.30	 &7.36	 &7.30	 &7.28	 &7.26\\
				  & 15648.510 	 &7.16	 &7.25 	 &7.27	 &7.24	 &7.37	 &7.26	 &7.20	 &...	 &7.35	 &7.32	 &7.17	 &... \\
				  & 15964.867	 &7.29	 &7.34	 &7.34	 &7.27	 &7.33	 &7.32	 &7.32	 &7.26	 &7.32	 &7.28	 &7.34	 &7.26\\
				  & 16040.657	 &7.29	 &7.25	 &7.21	 &7.14	 &7.30	 &7.18	 &7.35	 &7.23	 &7.27	 &7.22	 &7.30	 &7.28\\
				  & 16153.247	 &7.32	 &7.26	 &7.24	 &7.22	 &7.32	 &7.23	 &7.28	 &7.29	 &7.39	 &7.24	 &7.22	 &7.14\\
				  & 16165.032	 &7.32	 &7.32	 &7.34	 &7.28	 &7.32	 &7.27	 &7.33	 &7.31	 &7.40	 &7.37	 &7.29	 &7.17\\

$\langle$A(Fe)$\rangle$ $\pm$ $\sigma$ & & 7.29 $\pm$ 0.07&	7.29 $\pm$ 0.04&	7.26 $\pm$ 0.04&	7.24 $\pm$ 0.06&	7.32 $\pm$ 0.04&	7.26 $\pm$ 0.05&	7.32 $\pm$ 0.06&	7.26 $\pm$ 0.07&	7.36 $\pm$ 0.05&	7.31 $\pm$ 0.05&	7.28 $\pm$ 0.05&	7.23 $\pm$ 0.06\\
 & & & & & & & & & & & &\\
{\bf Na I}  	 & 16373.853 &6.16	 &6.05	 &6.02	 &6.10	 &6.16	 &6.08	 &6.05	 &...	 &6.14	 &...	 &6.19	 &6.10\\
                 & 16388.858 &6.11	 &6.05	 &6.10	 &6.11	 &6.14	 &6.11	 &6.18	 &6.06	 &6.18	 &6.13	 &6.10	 &6.13\\
$\langle$A(Na)$\rangle$ $\pm$ $\sigma$ & & 6.14 $\pm$ 0.03&	6.05 $\pm$ 0.01&	6.06 $\pm$ 0.04&	6.11 $\pm$ 0.01&	6.15 $\pm$ 0.01&	6.10 $\pm$ 0.02&	6.12 $\pm$ 0.07&	6.06 $\pm$ 0.01&	6.16  $\pm$ 0.02&	6.13 $\pm$ 0.01&	6.15 $\pm$ 0.05&	6.12 $\pm$ 0.02\\
 & & & & & & & & & & & &\\
{\bf Mg I} 	     & 15740.716 &7.21	 &7.25	 &7.26	 &7.20	 &7.36	 &7.20	 &7.23	 &7.18	 &7.35	 &7.29	 &7.19	 &7.12\\
                 & 15748.9   &7.31	 &7.33	 &7.35	 &7.30	 &7.37	 &7.25	 &7.35	 &7.23	 &7.41	 &7.35	 &7.31	 &7.28\\
                 & 15765.8   &7.32	 &7.28	 &7.21	 &7.26	 &7.31	 &7.26	 &7.37	 &7.17	 &7.37	 &7.30	 &7.29	 &7.29\\
                 & 15879.5   &7.31	 &7.32	 &7.29	 &7.34	 &7.31	 &7.31	 &7.32	 &7.28	 &7.32	 &7.30	 &7.26	 &7.25\\
                 & 15886.2   &7.43	 &7.45	 &7.42	 &7.45	 &7.50	 &7.49	 &7.45	 &7.47	 &7.50	 &7.47	 &7.50	 &7.50\\
                 & 15954.477 &7.40	 &7.39	 &7.44	 &7.44	 &7.42	 &7.44	 &7.48	 &7.44	 &7.49	 &7.42	 &7.43	 &7.41\\
$\langle$A(Mg)$\rangle$ $\pm$ $\sigma$ & & 7.33 $\pm$ 0.07&	7.34 $\pm$ 0.06&	7.33 $\pm$ 0.08&	7.33 $\pm$ 0.09&	7.38 $\pm$ 0.06&	7.33 $\pm$ 0.11&	7.37 $\pm$ 0.08&	7.30 $\pm$ 0.12&	7.41  $\pm$ 0.07&	7.36 $\pm$ 0.07&	7.33 $\pm$ 0.10&	7.31 $\pm$ 0.12\\
 & & & & & & & & & & & &\\
  
{\bf Al I}  	 & 16718.957 &6.33	 &6.31	 &6.38	 &6.30	 &6.36	 &6.27	 &6.37	 &6.26	 &6.42	 &6.33	 &6.27	 &6.27\\
                 & 16763.359 &6.35	 &6.34	 &6.26	 &6.26	 &6.33	 &6.28	 &6.29	 &6.21	 &6.34	 &6.28	 &6.27	 &6.22\\
$\langle$A(Al)$\rangle$ $\pm$ $\sigma$ & & 6.34 $\pm$ 0.02&	6.33 $\pm$ 0.02&	6.32 $\pm$ 0.06&	6.28 $\pm$ 0.02&	6.35 $\pm$ 0.02&	6.28 $\pm$ 0.01&	6.33 $\pm$ 0.04&	6.24 $\pm$ 0.03&	6.38 $\pm$ 0.04&	6.31 $\pm$ 0.03&	6.27 $\pm$ 0.01&	6.25 $\pm$ 0.03\\
 & & & & & & & & & & & &\\
 
{\bf Si I}   	 & 15361.161 &7.42	 &7.34	 &7.28	 &7.38	 &7.40	 &7.43	 &7.45	 &7.46	 &7.40	 &7.41	 &7.37	 &7.40\\	
                 & 15376.831 &7.46	 &7.44	 &7.55	 &7.43	 &7.51	 &7.44	 &7.44	 &7.53	 &7.53	 &7.48	 &7.49	 &7.45\\
                 & 15960.063 &7.44	 &7.43	 &7.46	 &7.37	 &7.52	 &7.37	 &7.46	 &7.36	 &7.51	 &7.49	 &7.43	 &7.31\\	
                 & 16060.009 &7.31	 &7.35	 &7.28	 &7.29	 &7.38	 &7.24	 &7.30	 &7.34	 &7.37	 &7.30	 &7.27	 &7.21\\
                 & 16094.787 &7.34	 &7.29	 &7.29	 &7.28	 &7.36	 &7.28	 &7.34	 &7.29	 &7.42	 &7.31	 &7.33	 &7.33\\
                 & 16215.670 &7.45	 &7.42	 &7.31	 &7.30	 &7.47	 &7.35	 &7.43	 &7.41	 &7.52	 &7.48	 &7.41	 &7.43\\
                 & 16680.770 &7.36	 &7.27	 &7.26	 &...	 &...	 &...	 &7.26	 &...	 &7.31	 &7.29	 &7.32	 &7.23\\
                 & 16828.159 &7.37	 &7.38	 &7.38	 &7.38	 &7.36	 &7.34	 &7.40	 &7.39	 &7.44	 &7.40	 &7.37	 &7.37\\            	
$\langle$A(Si)$\rangle$ $\pm$ $\sigma$ & & 7.39 $\pm$ 0.05&	7.37 $\pm$ 0.06&	7.35 $\pm$ 0.10&	7.35 $\pm$ 0.05&	7.43 $\pm$ 0.06&	7.35 $\pm$ 0.07&	7.39 $\pm$ 0.07&	7.40 $\pm$ 0.07&	7.44 $\pm$ 0.08&	7.40 $\pm$ 0.08&	7.37 $\pm$ 0.06&	7.34 $\pm$ 0.08\\
 & & & & & & & & & & & &\\

{\bf K I}	     & 15163.067 &4.94	 &4.92	 &4.93	 &4.87	 &5.01	 &4.90	 &4.94	 &4.78	 &4.94	 &4.96	 &4.87	 &4.86\\
				 & 15168.376 &4.89	 &4.90	 &4.91	 &4.89	 &4.81	 &4.91	 &4.91	 &4.73	 &4.94	 &4.93	 &4.89	 &4.89\\ 
$\langle$A(K)$\rangle$ $\pm$ $\sigma$ & & 4.92 $\pm$ 0.03&	4.91 $\pm$ 0.01&	4.92 $\pm$ 0.01&	4.88 $\pm$ 0.01&	4.91 $\pm$ 0.10&	4.91 $\pm$ 0.01&	4.93 $\pm$ 0.02&	4.76 $\pm$ 0.03&	4.94 $\pm$ 0.01&	4.95 $\pm$ 0.02&	4.88 $\pm$ 0.01&	4.88$\pm$ 0.02\\

 & & & & & & & & & & & &\\

{\bf Ca I}	 	 & 16136.823 &6.13	 &6.07	 &6.11	 &6.09	 &6.14	 &6.12	 &6.14	 &6.04	 &6.13	 &6.07	 &6.10	 &6.07\\
			 	 & 16150.763 &6.18	 &6.14	 &6.12	 &6.10	 &6.15	 &6.13	 &6.17	 &6.07	 &6.21	 &6.12	 &6.10	 &6.06\\
			 	 & 16155.236 &6.15	 &6.17	 &6.14	 &6.15	 &6.13	 &6.19	 &6.17	 &6.14	 &6.25	 &6.19	 &6.25	 &6.19\\
			 	 & 16157.364 &6.19	 &6.20	 &6.22	 &6.18	 &6.18	 &6.18	 &6.20	 &6.13	 &6.18	 &6.20	 &6.19	 &6.14\\
$\langle$A(Ca)$\rangle$ $\pm$ $\sigma$ & & 6.16 $\pm$ 0.02&	6.15 $\pm$ 0.05&	6.15 $\pm$ 0.04&	6.13 $\pm$ 0.04&	6.15 $\pm$ 0.02&	6.16 $\pm$ 0.03&	6.17 $\pm$ 0.02&	6.10 $\pm$ 0.04&	6.19 $\pm$ 0.04&	6.15 $\pm$ 0.05&	6.16 $\pm$ 0.06&	6.12 $\pm$ 0.05\\
 & & & & & & & & & & & &\\

{\bf Ti I}		 & 15543.756 &4.78	 &4.70	 &4.88	 &4.70	 &4.76	 &4.71	 &4.79	 &4.67	 &4.78	 &4.79	 &4.78	 &4.69\\ 
				 & 15602.842 &4.88	 &4.75	 &4.84	 &4.82	 &4.72	 &4.73	 &4.84	 &4.87	 &7.79	 &4.71	 &4.76	 &4.67\\
				 & 15698.979 &...	 &...	 &4.67	 &...	 &...	 &...	 &...	 &...	 &...	 &...	 &...	 &... \\
				 & 15715.573 &4.75	 &4.68	 &4.73	 &4.75	 &4.63	 &4.69	 &4.70	 &...	 &4.76	 &4.67	 &4.72	 &4.71\\
				 & 16635.161 &4.83	 &4.76	 &4.87	 &4.74	 &4.76	 &4.81	 &4.85	 &4.80	 &4.80	 &4.78	 &4.67	 &4.75\\
$\langle$A(Ti)$\rangle$ $\pm$ $\sigma$ & & 4.81 $\pm$ 0.05&	4.72 $\pm$ 0.03&	4.80 $\pm$ 0.08&	4.75 $\pm$ 0.05&	4.72 $\pm$ 0.05&	4.74 $\pm$ 0.05&	4.80 $\pm$ 0.06&	4.78 $\pm$ 0.09&	4.78 $\pm$ 0.01&	4.74 $\pm$ 0.05&	4.74 $\pm$ 0.04&	4.71 $\pm$ 0.03\\
 & & & & & & & & & & & &\\

{\bf V I} 		 & 15924.769 &3.97	 &3.90	 &3.86	 &3.88	 &...	 &3.87	 &3.85	 &3.82	 &3.82	 &3.79	 &3.91	 &3.84\\
$\langle$A(V)$\rangle$ $\pm$ $\sigma$ & & 3.97 $\pm$ ...&	3.90 $\pm$ ...&	3.86 $\pm$ ...& 3.88 $\pm$ ...&	... $\pm$ ...&	3.87 $\pm$ ...&	3.85 $\pm$ ...& 3.82 $\pm$ ...&	3.82 $\pm$ ...&	3.79 $\pm$ ...&	3.91 $\pm$ ...&	3.84 $\pm$ ... \\
 & & & & & & & & & & & &\\
  
{\bf Cr I} 	 	 & 15680.063  &5.46	 &5.38	 &5.42	 &5.41	 &5.46	 &5.40	 &5.46	 &5.50	 &5.48	 &5.41	 &5.43	 &5.42\\
$\langle$A(Cr)$\rangle$ $\pm$ $\sigma$ & & 5.46 $\pm$ ...&	5.38 $\pm$ ...&	5.42 $\pm$ ...&	5.41 $\pm$ ...&	5.46 $\pm$ ...&	5.40 $\pm$ ...&	5.46 $\pm$ ...&	5.50 $\pm$ ...&	5.48 $\pm$ ...&	5.41 $\pm$ ...&	5.43 $\pm$ ...&	5.42 $\pm$ ... \\

 & & & & & & & & & & & &\\

{\bf Mn I}       & 15159.0	  &5.18	 &5.11	 &5.17	 &5.17	 &5.17	 &5.17	 &5.18	 &5.18	 &5.21	 &5.21	 &5.18	 &5.10\\
		      	 & 15217.0	  &5.14	 &5.14	 &5.17	 &5.16	 &5.18	 &5.19	 &5.19	 &5.09	 &5.22	 &5.17	 &5.18	 &5.15\\
		     	 & 15262.0	  &5.23	 &5.17	 &5.18	 &5.15	 &5.19	 &5.21	 &5.19	 &5.12	 &5.23	 &5.20	 &5.14	 &5.13\\
$\langle$A(Mn)$\rangle$ $\pm$ $\sigma$ & & 5.18 $\pm$ 0.04&	5.14 $\pm$ 0.02&	5.17 $\pm$ 0.01&	5.16 $\pm$ 0.01&	5.18 $\pm$ 0.01&	5.19 $\pm$ 0.02&	5.19 $\pm$ 0.01&	5.13 $\pm$ 0.04&	5.22 $\pm$ 0.01&	5.19 $\pm$ 0.02&	5.17 $\pm$ 0.02&	5.13 $\pm$ 0.02\\
 & & & & & & & & & & & &\\

 {\bf Co I} 	 & 16757.7   &4.75	 &4.58	 &4.66	 &4.60	 &4.60	 &4.69	 &4.68	 &4.70	 &4.66	 &4.69	 &4.62	 &4.63\\
$\langle$A(Co)$\rangle$ $\pm$ $\sigma$ & & 4.75 $\pm$ ...&	4.58 $\pm$ ...&	4.66 $\pm$ ... &	4.60 $\pm$ ... &	4.60 $\pm$ ...&	4.69 $\pm$ ...&	4.68 $\pm$ ...&	4.70 $\pm$ ...&	4.66 $\pm$ ...&	4.69 $\pm$ ...&	46279 $\pm$ ...&	4.63 $\pm$ ...\\
 & & & & & & & & & & & &\\

{\bf Ni I}  	 & 15605.68	 &5.99	 &6.01	 &5.95	 &6.02	 &6.11	 &5.99	 &6.07	 &5.97	 &6.11	 &6.04	 &6.03	 &6.00\\
		  		 & 15632.654 &6.05	 &6.04	 &6.02	 &6.03	 &6.05	 &6.04	 &6.02	 &6.02	 &6.00	 &6.06	 &6.02	 &6.05\\
			  	 & 16584.439 &5.99	 &5.98	 &5.96	 &6.03	 &6.01	 &6.03	 &6.03	 &6.00	 &6.03	 &6.02	 &6.01	 &6.03\\
			  	 & 16589.295 &5.96	 &6.01	 &5.95	 &6.08	 &5.99	 &6.05	 &6.09	 &5.97	 &6.03	 &6.01	 &6.04	 &6.03\\
			  	 & 16673.711 &5.99	 &6.00	 &5.92	 &6.01	 &6.06	 &6.02	 &6.00	 &5.95	 &6.03	 &6.02	 &5.95	 &5.99\\
			  	 & 16815.471 &6.05	 &5.99	 &5.91	 &5.99	 &6.01	 &6.07	 &6.01	 &6.02	 &6.12	 &6.06	 &6.00	 &6.00\\
			  	 & 16818.76	 &6.04 	 &6.11	 &6.04	 &6.07	 &6.12	 &6.07	 &6.03	 &6.01	 &6.10	 &6.06	 &6.05	 &6.09\\
                 
$\langle$A(Ni)$\rangle$ $\pm$ $\sigma$ & & 6.01 $\pm$ 0.03&	6.02 $\pm$ 0.04&	5.96 $\pm$ 0.04& 6.03 $\pm$ 0.03&	6.04 $\pm$ 0.03&	6.04 $\pm$ 0.03&	6.04 $\pm$ 0.03&	5.99 $\pm$ 0.03&	6.04 $\pm$ 0.03&	6.04 $\pm$ 0.02&	6.01 $\pm$ 0.03&	6.03 $\pm$ 0.03\\
 & & & & & & & & & & & &\\
\enddata                                                                          
\end{deluxetable}

\clearpage
\begin{deluxetable}{ccccccc}
\tabletypesize{\scriptsize}
\tablecaption{Abundance Sensitivities}
\tablewidth{0pt}
\tablehead{
\colhead{Element} &
\colhead{$\Delta$T} &
\colhead{$\Delta$G} &
\colhead{$\Delta$$\rm \xi$}&
\colhead{$\Delta$M} &
\colhead{$\sigma$}\\
\colhead{} &
\colhead{(+50 K)} &
\colhead{(+0.20 dex)} &
\colhead{(+0.20 Km/s)} &
\colhead{(+0.20 dex)} &
\colhead{                           } 
}
\startdata
C  & +0.02	&	+0.02&	-0.03	&	+0.04	&	0.057\\
N  & -0.02	&	+0.02&	+0.00	&	+0.08	&	0.085\\
O  & +0.03	&	-0.03&	-0.06	&	+0.11	&	0.132\\
Na & +0.02	&	-0.02&	+0.00	&	+0.02	&	0.035\\
Mg & +0.02	&	-0.02&	+0.00	&	+0.04	&	0.049\\
Al & +0.05	&	-0.02&	-0.04	&	+0.04	&	0.078\\
Si & +0.00	&	-0.01&	-0.02	&	+0.05	&	0.055\\
K  & +0.03	&	-0.04&	-0.02	&	+0.01	&	0.055\\
Ca & +0.04	&	-0.02&	-0.02	&	+0.02	&	0.053\\
Ti & +0.09	&	+0.00&	-0.01	&	+0.05	&	0.103\\
V  & +0.04	&	+0.00&	-0.03	&	+0.03	&	0.058\\
Cr & +0.03	&	-0.02&	-0.03	&	+0.02	&	0.051\\
Mn & +0.02	&	+0.02&	-0.01	&	+0.01	&	0.032\\
Fe & +0.01	&	-0.02&	-0.05	&	+0.03	&	0.062\\
Co & +0.02	&	+0.00&	-0.05	&	+0.04	&	0.067\\
Ni & +0.00	&	+0.01&	-0.03	&	+0.03	&	0.044\\
 
\tablewidth{0pt}	

\enddata

\end{deluxetable}



\clearpage
\begin{deluxetable}{ccccc}
\tabletypesize{\scriptsize}
\tablecaption{Mean Abundances for NGC 2420}
\tablewidth{0pt}
\tablehead{
\colhead{Element} &
\colhead{$\langle$A(x)$\rangle$} &
\colhead{$\langle$[x/H]$\rangle$} &
\colhead{$\sigma$$_{\rm x}$ (dex)} &
\colhead{Bovy (2016): 68$\%$ Limits on} \\
\colhead{ } &
\colhead{ } &
\colhead{ } &
\colhead{ } &
\colhead{Abundance Scatter (dex)}
}
\startdata
   C    &    8.16    &	-0.23	&    0.04   &   0.03 \\
   N    &    7.79    &	+0.01	&    0.03   &   0.03 \\
   O    &    8.60    &	-0.06	&    0.03   &   0.06 \\
   Na   &    6.11    &	-0.06	&    0.04   &   0.06 \\
   Mg   &    7.34    &	-0.19	&    0.03   &   0.02 \\
   Al   &    6.31    &	-0.06	&    0.04   &   0.02 \\
   Si   &    7.38    &	-0.13	&    0.03   &   0.04 \\
   K    &    4.90    &	-0.18	&    0.05   &   0.06 \\
   Ca   &    6.15    &	-0.16	&    0.02   &   0.02 \\
   Ti   &    4.76    &	-0.14	&    0.03   &   0.05 \\
   V    &    3.86    &	-0.14	&    0.05   &   0.05 \\
   Cr   &    5.44    &	-0.20   &    0.04   &   ...  \\
   Mn   &    5.17    &	-0.22	&    0.03   &   0.03 \\
   Fe   &    7.29    &	-0.16	&    0.04   &   0.02 \\
   Co   &    4.66    &	-0.26	&    0.05   &   ... \\   
   Ni   &    6.02    &	-0.21	&    0.02   &   0.04 \\

\enddata
\end{deluxetable}

\tablerefs{}

\end{document}